%% file: main.tex
\documentclass[sigconf, nonacm]{acmart}
\AtBeginDocument{%
  }

\setcopyright{acmlicensed}
\copyrightyear{2027}
\acmYear{2027}
\acmDOI{XXXXXXX.XXXXXXX}

\acmConference[CHI '27]{CHI Conference on Human Factors in Computing Systems}{May 10--May 15, 2027}{Pittsburgh, USA}

\usepackage{multirow}
\usepackage{tabularx}
\usepackage{placeins}
\usepackage{enumitem}  
\usepackage{algorithmic}
\usepackage{algorithm}
\usepackage{float}
\usepackage[most]{tcolorbox}
\usepackage{ltablex}
\usepackage{stfloats}

\begin{document}

\title[TraceMind]{TraceMind: Predicting User Information Uptake from Low-Cost Interaction Traces during Human–LLM Content Co-Generation}

%
\author{Yu Mei}
\affiliation{
  \institution{Tsinghua University}
  \city{Beijing}
  \country{China}
}
\email{meiy24@mails.tsinghua.edu.cn}

\author{Fengyou Zu}
\affiliation{
  \institution{Tsinghua University}
  \city{Beijing}
  \country{China}
}
\email{zufy24@mails.tsinghua.edu.cn}

\author{Ruiwen Zhang}
\affiliation{
  \institution{Tsinghua University}
  \city{Beijing}
  \country{China}
}
\email{zrw22@mails.tsinghua.edu.cn}

\author{Jie Cai}
\affiliation{
  \institution{Nankai University}
  \city{Tianjin}
  \country{China}
}
\email{jie.cai1@outlook.com}

\author{Chang Liu}
\affiliation{
  \institution{University of Cambridge}
  \city{Cambridge}
  \country{United Kingdom}
}
\email{cl2123@cam.ac.uk}

\author{Zhoutong Ye}
\affiliation{
  \institution{Tsinghua University}
  \city{Beijing}
  \country{China}
}
\email{yezt24@mails.tsinghua.edu.cn}

\author{Chun Yu}
\affiliation{
  \institution{Tsinghua University}
  \city{Beijing}
  \country{China}
}
\email{chunyu@tsinghua.edu.cn}

\author{Yuanchun Shi}
\affiliation{
  \institution{Tsinghua University}
  \city{Beijing}
  \country{China}
}
\email{shiyc@tsinghua.edu.cn}

\renewcommand{\shortauthors}{Mei et al.}

\begin{abstract}
\input{data/abstract}
\end{abstract}

\ccsdesc[500]{Human-centered computing~User models}
\keywords{Information Uptake, AI-Assisted Writing, Large Language Models, User modeling, Interaction traces}
\begin{teaserfigure}
  \centering
  \includegraphics[width=1.0\textwidth]{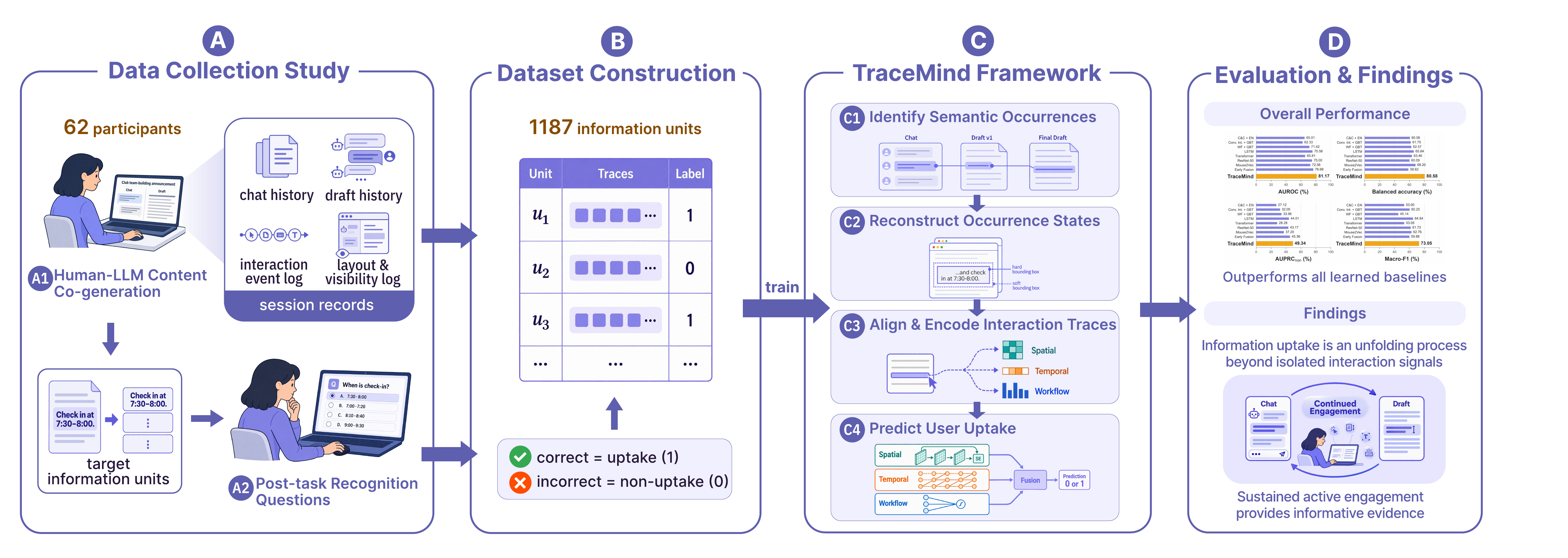}
  \caption{Overview of our study and TraceMind. (A) We collected human-LLM co-generation sessions from 62 participants and assessed dynamically extracted atomic information units using post-task recognition questions. (B) This yielded 1187 unit-level uptake labels paired with interaction traces. (C) TraceMind tracks each unit across Chat and Draft, aligns its interaction traces with changing on-screen states, and models spatial, temporal, and workflow-informed evidence to predict uptake. (D) TraceMind outperforms all learned baselines across four metrics, while our analyses show that information uptake unfolds across the interaction process, with sustained active engagement providing informative evidence beyond isolated interaction signals.}
  \Description{}
  \label{fig:teaser}
\end{teaserfigure}


\maketitle

\input{data/chap1}
\input{data/chap2}
\input{data/chap3}
\input{data/chap4}
\input{data/chap5}
\input{data/chap6}
\input{data/chap7}
\input{data/chap8}

\bibliographystyle{ACM-Reference-Format}
\bibliography{ref/main_citation}

\appendix
\input{data/appendix}

\end{document}

%% file: data/abstract.tex

In human-LLM content co-generation, AI-generated information can enter final artifacts without being adequately processed by users, creating risks when artifacts are shared or acted upon. We study whether recognition-level uptake of atomic information units can be assessed in open-ended co-generation and predicted from low-cost interaction traces. We collected data from 62 participants across three tasks. For each final draft, we extracted atomic information units and generated post-task recognition questions, yielding 1187 unit-level uptake labels. We present TraceMind, which tracks units across Chat and Draft histories, aligns interaction traces with changing on-screen layouts, and models spatial, temporal, and workflow-informed evidence. TraceMind outperformed all learned baselines across AUROC, $\mathrm{AUPRC}_{\mathrm{non}}$, balanced accuracy, and macro-F1. We found that uptake unfolds throughout interaction, with sustained active engagement providing informative evidence beyond isolated signals. Our work shifts human-LLM co-generation from content adoption toward what users actually take up, motivating uptake-aware systems grounded in low-cost interaction traces.

%% file: data/chap1.tex
\section{Introduction}
In human-LLM content co-generation, such as drafting an announcement or recruitment post, AI-generated information can enter final artifacts without being adequately processed by the user \cite{liu2026behavioral,yang2025modifying}. Overlooked information can lead to consequential errors or miscoordination, particularly in public settings such as journalism, public communication, or online posting \cite{nishal2026helping, fu2023comparing, jakesch2023co}. Avoiding such oversight requires modeling what users actually take up during human-LLM co-generation.

Prior work often focuses on how AI-generated content is adopted into final artifacts \cite{yang2025modifying, hwang202580}, offering limited insight into which information users have actually taken up. Information uptake can be considered at multiple levels, including recognition of information, comprehension of its meaning, and application in subsequent decisions or actions \cite{holderried2026impact, krathwohl2002revision}. In this work, we focus on \textbf{recognition-level information uptake}, operationalized as whether a user can correctly recognize what an atomic information unit in the final co-generated draft states immediately after completing the task \cite{freed2013effects, fruijtier2023identifying}. Following prior work on proposition segmentation and atomic evaluation, we define an \textbf{atomic information unit} as a self-contained statement that expresses a single piece of information \cite{hosseini2024scalable, min2023factscore}. 

In addition, prior work on modeling users' information-processing behavior has often relied on gaze tracking or other specialized sensing \cite{buscher2009you,hutt2022feasibility}, which can be difficult to deploy in everyday settings. In contrast, lower-burden mouse and keyboard interaction traces are readily available in everyday browser and editor interactions \cite{guo2012beyond,grinberg2018identifying}. However, whether such traces provide sufficient evidence of information uptake remains unclear.

In this paper, we ask whether such uptake can be assessed at the level of atomic information units in open-ended co-generation and predicted from low-cost interaction traces produced during the process.

Answering this question requires addressing two challenges. First, the \textbf{open-ended uptake assessment challenge}: unlike prior interaction-trace studies that assess predefined structural units (e.g., documents \cite{guo2012beyond}, paragraphs \cite{seifert2017focus}, individual HTML elements \cite{grusky2017modeling}, or predefined content blocks \cite{zhu2024modeling}), the information units in open-ended co-generation emerge dynamically and cannot be fixed as assessment targets in advance. Second, the \textbf{interaction-trace alignment challenge}: interaction events cannot be directly mapped to atomic information units, as the same information may move across Chat and Draft, change through revision, and be rendered differently over time.

To address the open-ended uptake assessment challenge, we collected data from 62 participants across three everyday co-generation tasks. For each final draft, we extracted atomic information units, retained those suitable for recognition assessment, and generated a post-task recognition question for each retained unit. Participants' answers provided performance-based uptake labels, yielding a dataset of 1187 labeled atomic information units paired with Chat and Draft histories, interaction events, and layout snapshots.

To address the interaction-trace alignment challenge, we present TraceMind, a framework that recovers and aligns the fragmented interaction history of each atomic information unit and uses the resulting traces to predict information uptake. TraceMind first identifies each atomic information unit's semantic occurrences across Chat and Draft histories by locating spans that express the same information. It then recovers the on-screen layouts corresponding to these occurrences and identifies the interaction records associated with them over time. Finally, TraceMind models the unit-aligned traces as complementary spatial, temporal, and workflow-informed evidence to predict uptake.

Evaluated on the collected dataset, TraceMind achieved 81.17\% AUROC, 80.58\% balanced accuracy, 49.34\% $\mathrm{AUPRC}_{\mathrm{non}}$, and 73.05\% macro-F1, outperforming all learned baselines across all four metrics. Our findings suggest that information uptake is an unfolding process that cannot be captured by isolated interaction signals alone, and that sustained active engagement (e.g., revisiting, revising, and cross-surface checking) provides particularly informative evidence for modeling uptake.

The contributions of this work are as follows:
\begin{itemize}
    \item \textbf{A performance-based method for assessing recognition-level information uptake in open-ended co-generation}, using dynamically derived atomic information units as assessment targets.
    \item \textbf{TraceMind, a framework for predicting information uptake from low-cost interaction traces}, by aligning users' evolving interactions with atomic information units across the co-generation process.
    \item \textbf{Empirical insights into information uptake as an unfolding process}, highlighting sustained active engagement as particularly informative for modeling uptake. Our findings motivate the design of uptake-aware human-LLM co-generation systems.
\end{itemize}

%% file: data/chap2.tex
\section{Related Work}
Our work draws on three areas of HCI research. First, we review human-LLM content co-generation as a broad HCI task context and highlight our shift from prior work's focus on the co-generated content itself to what users actually take up from that content. Second, we review subjective and performance-based approaches to user-state assessment, highlighting the challenge of extending performance-based measurement to open-ended tasks. Finally, we review interaction-trace-based user modeling, focusing on the opportunities and limitations of low-cost traces for modeling how users engage with evolving information.

\subsection{Human-LLM Content Co-Generation and Information Uptake}
Human-LLM content co-generation has become common across a wide range of HCI domains, including creative writing \cite{chung2022talebrush,yuan2022wordcraft}, educational writing \cite{jelson2026empirical}, professional communication \cite{fu2023comparing}, and personal journaling \cite{kim2024diarymate,zhou2025journalaide}. Across these settings, LLMs can support users in diverse roles, such as generating ideas, drafting text, and revising content \cite{gero2023social,ippolito2022creative,lee2022coauthor}. Prior work has also examined how users collaborate with LLMs during this process, such as how they accept suggestions, edit generated text, and divide writing work with AI \cite{dhillon2024shaping,bhat2026reactive,mysore2025prototypical}.

Recent studies have raised concerns about how users engage with AI-generated content during human-LLM co-generation. Work on content adoption examines whether and how AI-generated material enters the final artifact \cite{lee2022coauthor,hwang202580} and shows that users may substantially reuse AI suggestions \cite{a2026overreliance}. Recent work also examines users' overreliance on AI, showing that users may inappropriately accept incorrect AI outputs \cite{vasconcelos2023explanations,bo2025rely}. Studies have further found that users may copy or accept AI-generated content with limited scrutiny \cite{liu2026behavioral,bhat2026reactive}, while AI assistance can shape users' final expressions or opinions \cite{jakesch2023co,hoque2024hallmark,zindulka2026ai}. Together, this work highlights the risks of excessive reliance on AI-generated content, but largely focuses on the co-generated content itself---what users adopt, rely on, or ultimately produce.

We instead focus on what users themselves take up from co-generated content. Prior work has begun to reveal consequences of AI assistance for users' memory and awareness, including reduced recall of what users wrote \cite{kosmyna2025your} and confusion about whether content originated from themselves or AI \cite{zindulka2026ai}. Building on these observations, we seek to measure and predict recognition-level information uptake, operationalized as whether a user can correctly recognize what an atomic information unit in the final co-generated draft states immediately after completing the task \cite{freed2013effects,fruijtier2023identifying}. Our goal is to model which information users have actually taken up in co-generated content, enabling future co-generation systems to identify content that may require additional user attention or review.

\subsection{Assessing User States in HCI Tasks}
Subjective measures are widely used in HCI to assess user states such as perceived learning, confidence, trust, workload, and engagement. For example, search-as-learning studies collect perceived learning outcomes \cite{collins2016assessing}, AI-assisted decision-making studies use self-reported confidence to examine reliance and calibration \cite{ma2024you}, and recent work measures perceived self-regulated learning during information seeking \cite{urgo2026questionnaire}. Workload is also commonly assessed through subjective scales such as NASA-TLX \cite{babaei2025should,lee2026nasa}. These measures are easy to collect, but they may not align with behavioral or performance outcomes and can be affected by retrospective recall and individual calibration \cite{urgo2026questionnaire,ma2024you,babaei2025should}. They also require users to explicitly report their state, which can add burden and interrupt natural interaction \cite{kang2022understanding}.

Performance-based assessment instead asks users to demonstrate what they recognize, recall, or understand after interaction. For example, search-as-learning research uses pre- and post-task knowledge tests \cite{collins2016assessing,gadiraju2018analyzing,urgo2022learning}; health communication studies use recognition and recall tests \cite{freed2013effects,pajor2020effects}; and visualization research evaluates recall and comprehension of presented information \cite{kim2017explaining,zdanovic2022influence}. These approaches provide directly scored outcomes rather than relying only on users' own judgments. Among these measures, recognition is commonly assessed through statement recognition or multiple-choice questions \cite{freed2013effects}. More broadly, automatic question-generation research has shown that assessment questions can be generated from source text \cite{kurdi2020systematic}, and recent work has explored LLM-generated multiple-choice questions for scalable assessment \cite{laupichler2024large,holzing2026fine}. Together, these studies provide a foundation for constructing performance-based recognition assessments from textual content.

However, many performance-based assessments rely on \textit{fixed, predefined assessment units}. For example, search studies define topics and knowledge prompts before the task \cite{collins2016assessing,gadiraju2018analyzing}; health and visualization studies assess users on fixed texts or datasets \cite{freed2013effects,kim2017explaining}; and related interaction studies organize measurement around predefined documents, paragraphs, or interface elements \cite{guo2012beyond,grusky2017modeling,seifert2017focus}. Open-ended human-LLM co-generation is different: each user produces a different artifact, and its information units emerge during interaction. Existing co-writing studies therefore tend to evaluate content adoption, artifact-level outcomes, or user perceptions \cite{lee2022coauthor,bhat2026reactive,hwang202580}, rather than performance on the specific information contained in each user's final artifact. Our work addresses this gap by extending performance-based uptake assessment to open-ended co-generation.

\subsection{Modeling User States from Interaction Traces}
Interaction traces have long been used as indirect evidence of user states, such as attention, engagement, and information processing. In web search, dwell time, scrolling, and cursor behavior have been used to estimate document relevance and user interest \cite{huang2011no,guo2012beyond}. In online reading, viewport exposure, reading depth, and backtracking have supported models of attention and engagement \cite{grusky2017modeling,smadja2019understanding}. Related work has also used interaction behavior to predict session-level knowledge gain \cite{yu2018predicting} and to characterize AI reliance and overreliance \cite{ma2023should,liu2026behavioral}. 

A major line of this work relies on gaze and other specialized sensing. Gaze-based studies use fixations, refixations, regressions, and reading paths to model reading behavior and user attention \cite{buscher2009you,gu2024skimmers}. Such signals provide rich spatial evidence, but introduce practical constraints: dedicated eye trackers, wearable devices, or continuous camera-based sensing can be difficult to deploy in everyday settings and may raise privacy concerns \cite{hutt2022feasibility,wang2026mind}. These constraints motivate greater use of interaction traces that are already available in ordinary interfaces.

Lower-cost approaches rely on browser- and editor-level traces such as dwell time, viewport exposure, scrolling, cursor movement, text selection, copying, and editing \cite{grinberg2018identifying,lee2016spotlights}. Because these traces are naturally produced during routine interaction, they provide opportunities for modeling user states in everyday interfaces \cite{guo2012beyond,grinberg2018identifying}. These traces have been used to identify engagement patterns \cite{grinberg2018identifying}, estimate document relevance \cite{guo2012beyond}, predict knowledge gain \cite{yu2018predicting}, and characterize behavioral indicators of AI overreliance \cite{liu2026behavioral}.

However, low-cost interaction traces remain indirect and ambiguous proxies for user states \cite{lin2023scanning,grinberg2018identifying}. Moreover, prior work often summarizes such traces within fixed analysis scopes, such as pages, documents, sessions, or task instances \cite{grinberg2018identifying,guo2012beyond,yu2018predicting,ma2024you}. This makes such approaches difficult to apply directly to open-ended co-generation. We instead follow how users' interactions with the same piece of information evolve over time, enabling information uptake to be modeled across the co-generation process.

%% file: data/chap3.tex
\section{Data Collection Study}
\label{section: DATA COLLECTION STUDY}
To obtain fine-grained uptake labels in open-ended co-generation, we conducted a controlled data collection study. Our platform dynamically derived atomic information units from each final draft and generated corresponding post-task recognition questions. Participants' responses yielded unit-level uptake labels, which were paired with the interaction traces recorded throughout co-generation to form the resulting dataset. We next describe the study design, data collection system, and resulting dataset.

\subsection{Study Design}
\subsubsection{Participants.} 
We recruited 62 participants through social media (30 identified as female and 32 as male; aged 18–30 years, $M = 21.48$, $SD = 1.95$). Participant demographics are provided in Appendix~\ref{appendix: Participant Demographics}. On 5-point Likert scales, participants' self-reported familiarity with conversational LLMs averaged 4.10 ($SD = 0.43$), and their frequency of LLM use over the past month averaged 4.19 ($SD = 0.84$; 1 = almost never and 5 = multiple times daily). For writing-related tasks (e.g., drafting announcements, essays, or reports) specifically, participants reported both the frequency (1 = never, 5 = always) and the degree of reliance (1 = entirely self-written, 5 = almost entirely LLM-dependent) on LLMs, averaging 4.23 ($SD = 0.68$) and 3.89 ($SD = 0.84$), respectively. Each participant received \$8 as compensation.

\begin{table*}[b]
\caption{Human-LLM content co-generation tasks in our data collection study. Key entities are anonymized (e.g., Historic Town H, Apartment A, and Scholar P1).}
\label{tab:data-collection-task-settings}
\small
\begin{tabular}{ll}
\hline
\multicolumn{1}{c}{\textbf{Task}}                                         & \multicolumn{1}{c}{\textbf{Settings}}                                                                                                                                                                                                                                                                                                                                                                                                                                    \\ \hline
\begin{tabular}[c]{@{}l@{}}Club team-building\\ announcement\end{tabular} & \begin{tabular}[c]{@{}l@{}}Organize an annual club team-building trip to Historic Town H, including coach transportation, visits\\ to several local attractions, and dinner at Restaurant R. Draft an announcement introducing the activity\\ purpose, itinerary, venues, transportation,  attire, local cuisine, preparation, and other useful information.\end{tabular}                                                                                                \\ \hline
\begin{tabular}[c]{@{}l@{}}Roommate recruit-\\ ment post\end{tabular}     & \begin{tabular}[c]{@{}l@{}}Recruit one roommate for a D2--4$\times$4 unit at Apartment A during a two-month summer program at\\ University B. The participant and two other students have already confirmed their stay. Draft a public\\ post introducing the apartment, amenities, neighborhood, transportation,  move-in arrangements,\\ current roommates, and expectations for shared living.\end{tabular}                                                           \\ \hline
\begin{tabular}[c]{@{}l@{}}Lecture announce-\\ ment\end{tabular}          & \begin{tabular}[c]{@{}l@{}}The university will host an AI lecture at 2:00 p.m. the following Friday, featuring three academicians\\ from University A---Scholar P1, Scholar P2, and Scholar P3. Each scholar will give an academic talk \\ followed by a Q\&A session. Draft an announcement introducing the speakers, their research areas and\\ representative work, the event theme and value, suggested preparation, and other useful information.\end{tabular} \\ \hline
\end{tabular}
\end{table*}

\subsubsection{Task Design}
We designed three co-generation tasks: a club team-building announcement, a roommate recruitment post, and a lecture announcement. Each participant was randomly assigned to one task. \autoref{tab:data-collection-task-settings} summarizes the overall task settings; Appendix~\ref{appendix: Participant-Facing Task Instructions} provides the complete participant-facing task instructions.

We selected the tasks based on four considerations. First, they reflect familiar everyday scenarios in which people plausibly use LLMs for writing, increasing ecological validity. Second, unnoticed errors in the resulting public-facing final draft could lead to serious real-world consequences, underscoring the importance of users’ information uptake. Third, each prompt allowed substantial flexibility in content creation, enabling diverse co-generation behaviors. Fourth, each deliverable contained multiple discrete details (e.g., schedules, locations, rules, profiles) that could be assessed through information-unit-level recognition questions.

\subsubsection{Procedure.} 
Participants first provided informed consent and completed a pre-task questionnaire. They then reviewed a study guide introducing the co-generation task, the TraceMind system, and the study procedure. Participants accessed our publicly hosted web-based study platform directly through their browsers to complete the co-generation task (see Section~\ref{section: User Interface}). To ensure reliable capture of mouse interaction traces, they were required to complete the study on a laptop using an external mouse. To ensure sufficient interaction and analyzable content, participants were required to co-generate a draft of at least 1500 characters. After completing the task, they uploaded their screen recordings. Each session lasted approximately 30 minutes.

\subsection{Data Collection System}

\begin{figure*}[htbp] 
\centering
\includegraphics[width=1.0\textwidth]{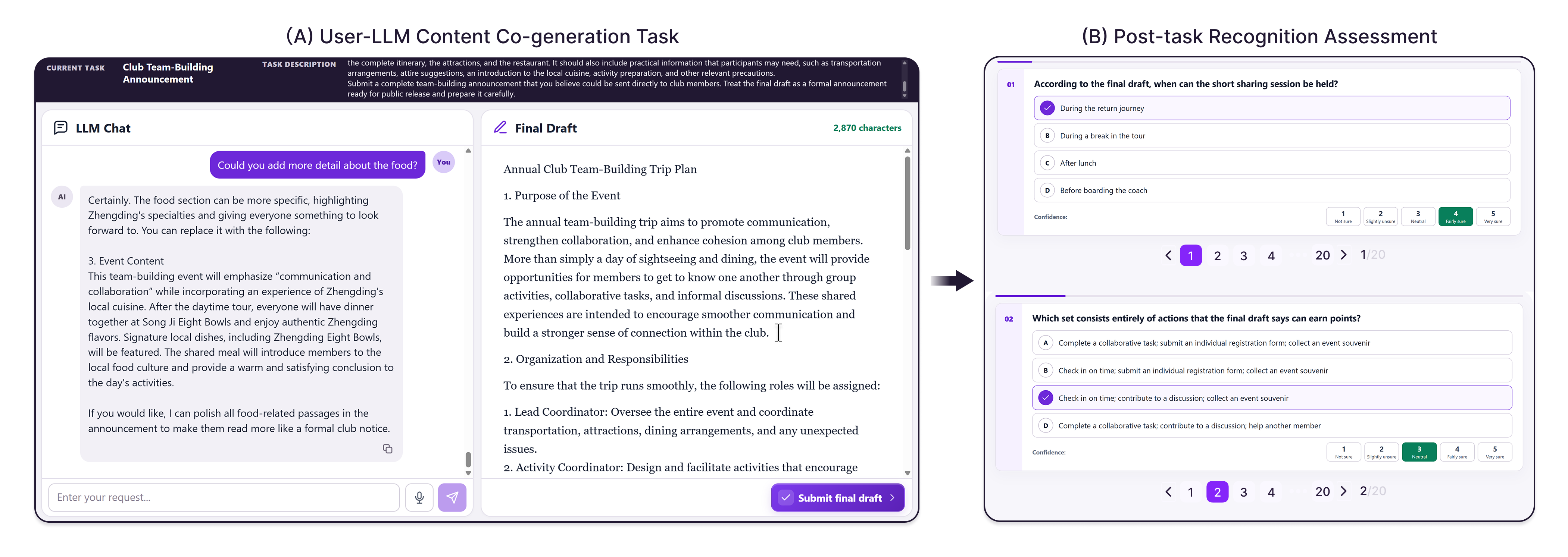}
\caption{Data collection study interfaces. (A) Participants co-generated content with an LLM Chat surface and an editable Draft surface. (B) After submitting the draft, they answered recognition questions and rated their confidence.}
\label{fig:UI}
\end{figure*}

\subsubsection{User Interface}
\label{section: User Interface}
The study platform comprised two work surfaces (\autoref{fig:UI}). During the co-generation task (\autoref{fig:UI}A), participants worked in a side-by-side workspace with the LLM chat on the left and an editable Draft panel on the right. Participants could compose prompts by typing or using voice dictation. They could copy content from LLM responses into the draft or write and revise the draft directly. The Chat surface used GPT-5.4 as the underlying LLM.

After submitting their final drafts, participants proceeded to the post-task recognition interface (\autoref{fig:UI}B). The interface presented four-option recognition questions based on the final draft. For each question, participants selected an answer and rated their confidence on a five-point scale before proceeding. Questions were presented one at a time without backward navigation to preserve item independence and limit cross-item cueing. We use answer correctness as the primary outcome label for information uptake. As a complementary analysis, we examine how self-reported confidence relates to recognition correctness and interaction traces in Section~\ref{section: Self-reported Confidence and Metacognitive Awareness of Information Uptake}.

\begin{table*}[htbp]
\caption{Interaction events recorded across the Chat and Draft surfaces during the data collection study.}
\label{tab:raw-behavior-events}
\small
    \begin{tabular}{@{}
        p{0.20\linewidth}
        p{0.25\linewidth}
        p{\dimexpr0.55\linewidth-4\tabcolsep\relax}
    @{}}
    \toprule
    \multicolumn{1}{c}{\textbf{Type}} &
    \multicolumn{1}{c}{\textbf{Event}} &
    \multicolumn{1}{c}{\textbf{Description}} \\
    \midrule
    \multirow{9}{=}{Shared across Chat and Draft surfaces}
    & \texttt{pointer\_move}
    & The user moves the pointer over content in either surface. \\
    \cmidrule(l){2-3}
    & \texttt{pointer\_click}
    & The user clicks content or an interactive element. \\
    \cmidrule(l){2-3}
    & \texttt{scroll\_change}
    & The user scrolls the Chat or Draft surface. \\
    \cmidrule(l){2-3}
    & \texttt{text\_selection\_episode}
    & The user selects a text span in a chat message or the Draft. \\
    \cmidrule(l){2-3}
    & \texttt{clipboard\_copy}
    & The user copies selected text from either surface. \\
    \cmidrule(l){2-3}
    & \texttt{clipboard\_cut}
    & The user cuts selected text from an editable field in either surface. \\
    \cmidrule(l){2-3}
    & \texttt{clipboard\_paste}
    & The user pastes text into the chat input or Draft. \\
    \cmidrule(l){2-3}
    & \texttt{focus\_in/out}
    & The user moves the interaction focus into or out of a surface element. \\
    \cmidrule(l){2-3}
    & \texttt{activity\_idle\_start/end}
    & The user remains inactive for at least 30 seconds and subsequently resumes interaction. \\
    \midrule
    \multirow{3}{=}{Chat surface-specific}
    & \texttt{chat\_submit}
    & The user submits a prompt composed through keyboard, voice, or mixed input. Keyboard interactions within the same prompt are represented collectively at submission. \\
    \cmidrule(l){2-3}
    & \texttt{voice\_input\_start}
    & The user activates the microphone and begins composing a prompt through voice input. \\
    \cmidrule(l){2-3}
    & \texttt{voice\_input\_stop}
    & The user stops voice input manually or reaches the recording time limit. \\
    \midrule
    \multirow{3}{=}{Draft surface-specific}
    & \texttt{text insertion}
    & The user inserts or revises draft text through keyboard or input-method input, including typing over selected text. \\
    \cmidrule(l){2-3}
    & \texttt{text deletion}
    & The user deletes draft text using Backspace, Delete, or an equivalent editing operation without inserting replacement text. \\
    \cmidrule(l){2-3}
    & \texttt{final\_submit}
    & The user submits the completed draft. \\
    \bottomrule
    \end{tabular}
\end{table*}

\subsubsection{Interaction Event Logging}
We instrumented the web-based study platform with JavaScript event listeners to record participants' interaction behaviors during co-generation. The logger sampled timestamped pointer coordinates $(x,y)$ at approximately 10~Hz and recorded user-initiated operations in both the Chat and Draft surfaces. As summarized in \autoref{tab:raw-behavior-events}, we distinguish user behaviors captured consistently across both surfaces from those specific to the Chat or Draft surface.

To align interaction events with the content displayed on screen, we recorded change-triggered snapshots of the Chat and Draft surfaces. When the surface content changed (e.g., when an LLM response appeared or the draft was edited), the logger captured paired content and layout snapshots, including the full surface text, block-level offsets, and line- and grapheme-level bounding boxes. When the visible viewport changed (e.g., during scrolling), the logger captured visibility snapshots containing the scroll position, viewport bounds, visible text lines, and the visible proportion of each line.

Together, these records preserve the Draft history, Chat history, interaction event log, and layout and visibility snapshots used as session record inputs to TraceMind (\autoref{fig:pipeline}).

\subsubsection{Post-task Question Generation}
\label{section: Post-task Question Generation}

\begin{figure*}[b] 
\centering
\includegraphics[width=1.0\textwidth]{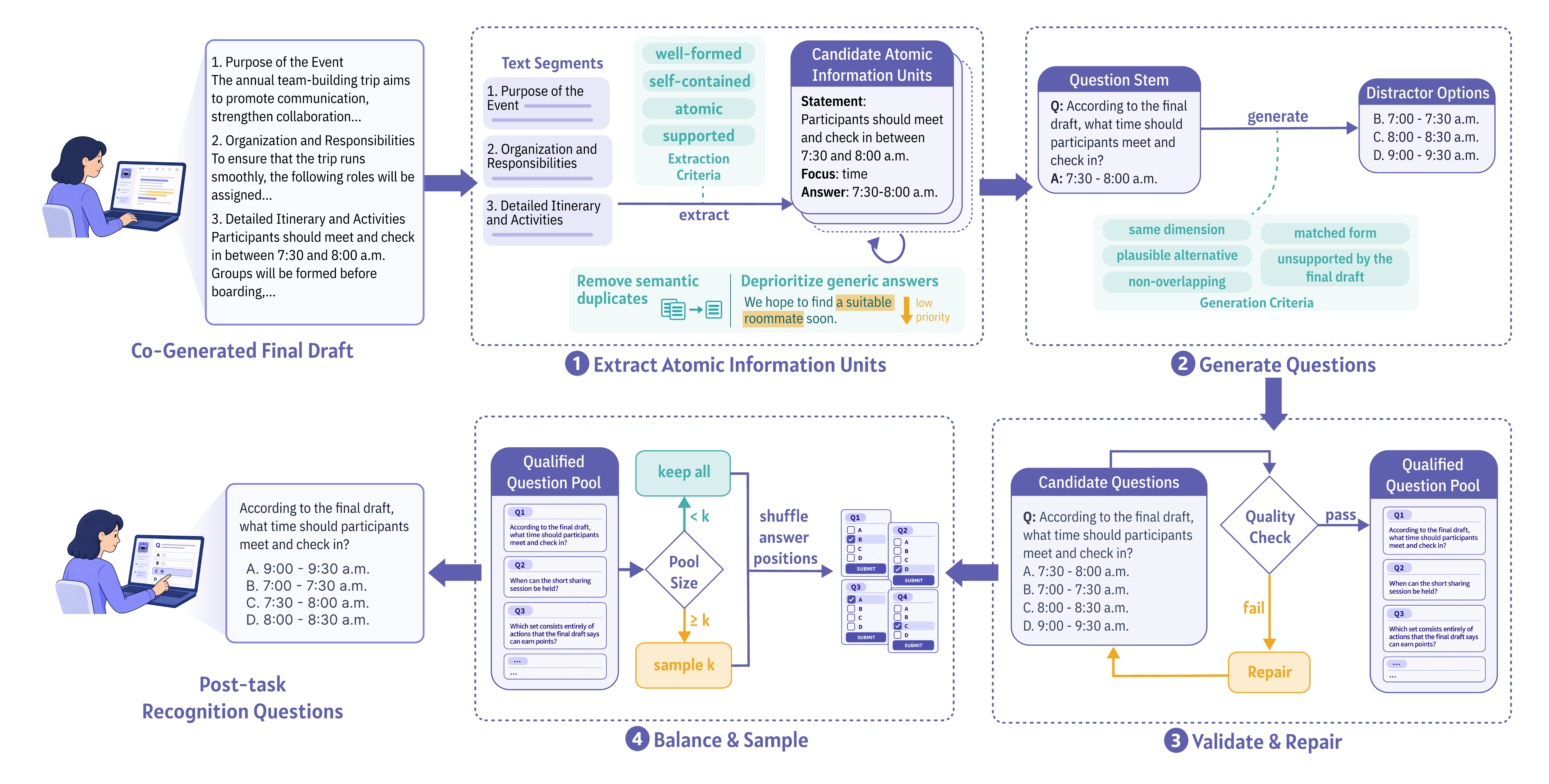}
\caption{Question generation pipeline. Given each final draft, the system (1) extracts and filters candidate atomic information units, (2) generates a recognition question with one correct answer and three distractors for each retained unit, (3) validates and repairs candidate questions, and (4) samples from the qualified pool and shuffles answer positions to produce the final post-task recognition questions.}
\label{fig:question-generation-pipeline}
\end{figure*}

After participants submitted their final drafts, the system automatically generated post-task recognition questions to assess participants' immediate information uptake (\autoref{fig:question-generation-pipeline}).

The system first divided each final draft into text segments and extracted candidate atomic information units using criteria adapted from prior work on proposition segmentation \cite{hosseini2024scalable}. Each unit had to (1) be well-formed, (2) be self-contained, (3) express a single piece of information, and (4) be supported by the draft. The system then removed semantically equivalent units (e.g., ``Check-in is scheduled for 7:30-8:00 a.m.'' and ``The check-in window is 7:30-8:00 a.m.''). Units likely to yield overly generic answers were identified and assigned lower priority during question sampling. For example, ``a suitable roommate'' in ``We hope to find a suitable roommate soon'' provides no specific roommate characteristics and therefore makes it difficult to generate discriminative distractors.

For each retained atomic information unit, the system generated a question stem and the correct answer, then generated three distractors using the complete final draft as context. Distractor generation followed five criteria: (1) same dimension: match the correct answer's semantic type and specificity; (2) plausible alternatives: be credible in the scenario and not easily ruled out by common knowledge; (3) non-overlapping: remain semantically distinct from all other options, without containment, overlapping ranges, or shared endpoints; (4) matched form: use the same language and comparable length, grammar, and formatting to avoid answer cues; and (5) unsupported by the final draft: no passage in the final draft may support the distractor as a valid answer to the question. For example, a roommate recruitment post might list a gym, study area, package collection service, and laundry room as apartment amenities. For a question asking which amenity is listed, ``a laundry room'' cannot serve as a distractor for ``a gym'', because both are valid answers.

Each resulting four-option, single-choice question item was validated for (1) clear and unambiguous wording, (2) a unique correct answer directly supported by the final draft, (3) comparable and non-overlapping options, and (4) dependence on the final draft rather than common knowledge. Invalid items underwent one targeted repair attempt and were discarded if they still failed validation.  

If the qualified question pool exceeded $K_{\max}=20$, the system performed stratified sampling based on the locations of the source text supporting each answer. Finally, given that LLM-generated questions can exhibit biases in the positions assigned to correct answers \cite{tang2026large}, the system then randomly shuffled the answer options to mitigate positional bias.

We used GPT-5.4 for information-unit extraction, question generation, and question validation throughout this pipeline.

\subsection{Dataset Construction}
\label{section: Dataset Construction}
We constructed the dataset from 62 participants, each contributing one completed study session. For each session, the archived data included the Chat and Draft histories, interaction-event logs, surface and visibility snapshots, recognition questions and responses, extracted information units, and their corresponding occurrences and exposure intervals.

The 62 sessions included 1187 information units with valid labels based on post-task recognition responses. These included 428 units from the club team-building announcement task, 423 from the roommate recruitment post task, and 336 from the lecture announcement task. Participants correctly recognized 881 units and incorrectly recognized 306, corresponding to an overall recognition accuracy of 74.22\%. On average, participants spent 21.64 minutes in the writing interaction phase ($SD = 7.03$) and 27.95 minutes in the full session including the post-task recognition assessment ($SD = 11.39$); each session contained 17.19 chat messages, 33.94 draft versions, and 3525.79 interaction events. Each session also contained dense surface-state logs, with an average of 8245.27 snapshots capturing changes in surface content, rendered layout, and visibility.

%% file: data/chap4.tex
\section{TraceMind Framework}

\begin{figure*}[b]
    \centering
    \includegraphics[width=1.0\linewidth]{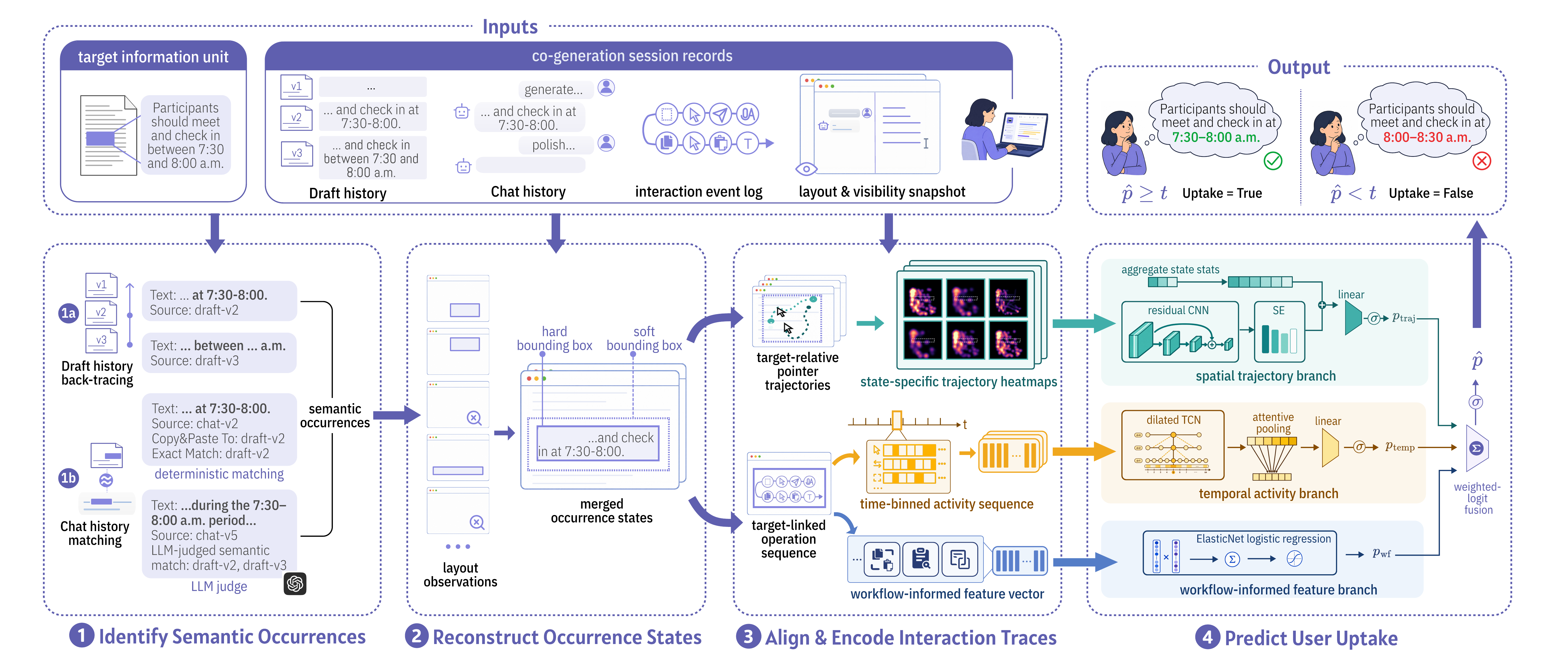}
    \caption{Overview of the TraceMind framework. Given a target information unit and logs from a user--LLM co-generation session, TraceMind (1) identifies the unit's semantic occurrences across the Draft and Chat histories, (2) reconstructs their layout-aware occurrence states, (3) aligns and encodes spatial, temporal, and workflow-informed interaction traces, and (4) combines three prediction branches to estimate information uptake.}
    \label{fig:pipeline}
\end{figure*}

In this section, we present the TraceMind framework (see \autoref{fig:pipeline}). TraceMind takes as input a target information unit from the final draft and records from the corresponding user--LLM co-generation session. These session records include the Draft history, Chat history, interaction event log, and layout and visibility snapshots. For each target information unit, TraceMind predicts whether the user took up that information. 

The key idea of TraceMind is to organize fragmented cross-interface interaction traces around where and how each information unit appeared over time. For each unit, TraceMind first identifies its semantic occurrences across the Chat and Draft histories (Section~\ref{section: Identifying Semantic Occurrences of Each Information Unit}) and reconstructs layout-aware occurrence states that capture how these occurrences were rendered and exposed over time (Section~\ref{section: Reconstructing Layout-Aware Occurrence States for Each Semantic Occurrence}). Next, it aligns and encodes the associated interaction traces into spatial, temporal, and workflow-informed representations (Section~\ref{section: Aligning and Encoding Interaction Traces}). Finally, TraceMind models these representations in three separate branches and combines their predictions into the final uptake prediction (Section~\ref{section: Information-Uptake Prediction Model}).

\subsection{Identifying Semantic Occurrences of Each Information Unit}
\label{section: Identifying Semantic Occurrences of Each Information Unit}
Throughout the co-generation process, the same information may first appear in the Chat, later be transferred to the Draft, and then be revised as the draft evolves. Therefore, given a target information unit in the final draft, TraceMind first identifies corresponding spans across the Draft and Chat histories that express the same information. We refer to each identified span as a \textbf{semantic occurrence} of the target information unit. As shown in Step 1 of \autoref{fig:pipeline}, TraceMind identifies these occurrences in two steps: Draft history back-tracing and Chat history matching.

\textbf{Draft history back-tracing.} We begin with the supporting span of the target unit in the final draft, associated with the unit's post-task recognition question (Section~\ref{section: Post-task Question Generation}). This span anchors the target unit to its occurrence in the final draft. Using text correspondence and recorded provenance, TraceMind traces the span backward through the Draft history to recover earlier spans in which the same information was retained or revised. Each recovered span is linked to the Draft version in which it appeared.

\textbf{Chat history matching.} TraceMind next searches the Chat history for spans corresponding to the recovered Draft occurrences. It first accepts deterministic matches supported by exact text correspondence or recorded copy-and-paste provenance. For candidates that cannot be resolved deterministically, an LLM judge determines whether a Chat span expresses the same information despite different wording and, if so, returns the matched span. We use GPT-5.4 as the LLM judge. Together, the accepted Draft and Chat spans form the semantic occurrences of the target information unit, which are then passed to layout-aware state reconstruction (Section~\ref{section: Reconstructing Layout-Aware Occurrence States for Each Semantic Occurrence}).

\subsection{Reconstructing Layout-Aware Occurrence States for Each Semantic Occurrence}
\label{section: Reconstructing Layout-Aware Occurrence States for Each Semantic Occurrence}

\begin{figure*}[b]
    \centering
    \includegraphics[width=0.6\textwidth]{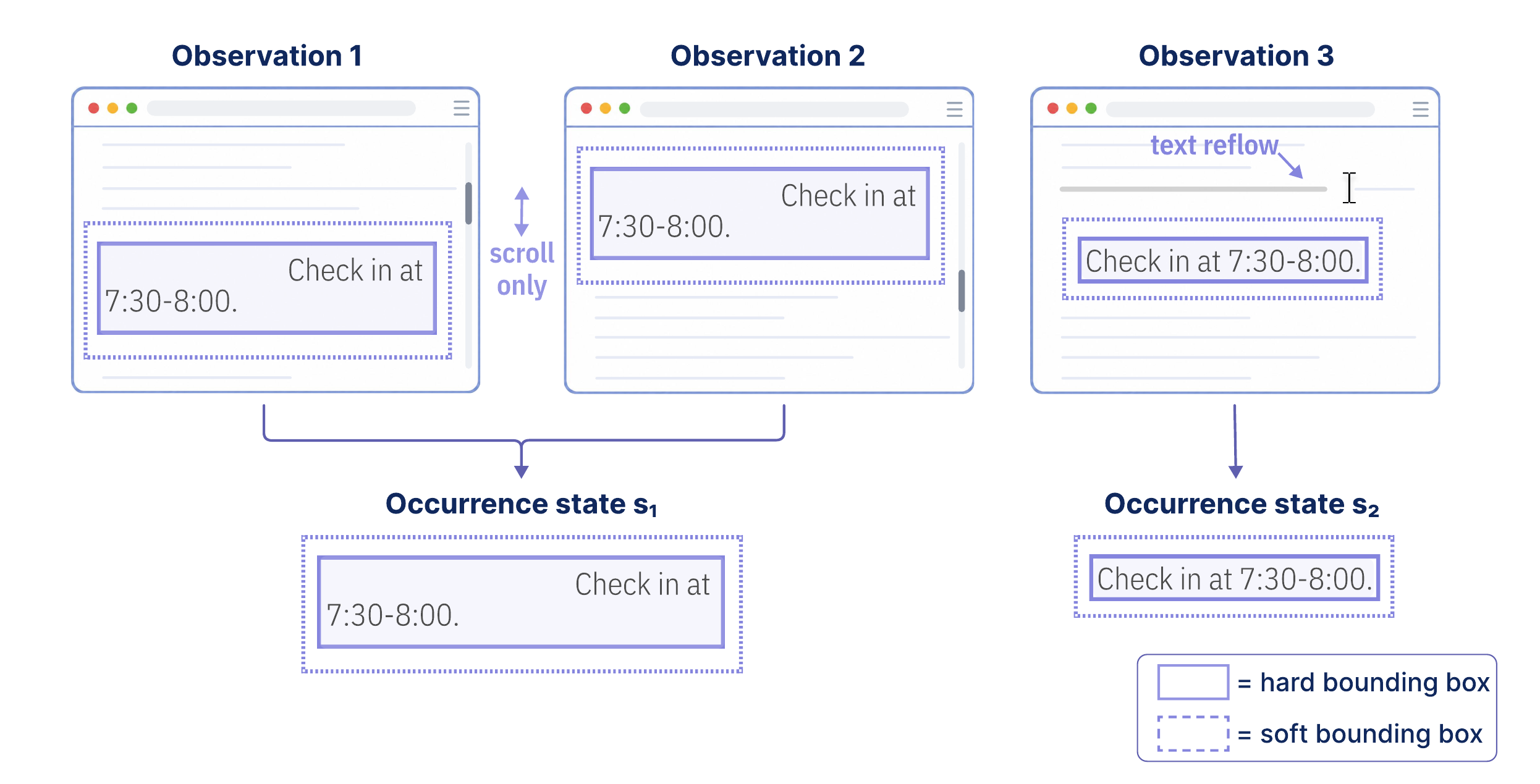}
    \caption{Reconstructing layout-aware occurrence states from layout observations. Observations are assigned to the same state only when they share the same interface context, text content, visibility condition, and approximately the same target bounding-box shape. For example, Observations 1 and 2 differ only in scroll position and are therefore merged into state $s_1$. In contrast, text reflow changes the shape in Observation 3, forming state $s_2$.}
    \label{fig:occurrence-states}
\end{figure*}

The same semantic occurrence may be rendered under different spatial conditions as the Chat and Draft evolve and the user scrolls. Because spatial interaction traces are meaningful only relative to the layout in which they were recorded, TraceMind represents each distinct rendering condition of a semantic occurrence as an \textbf{occurrence state}.

As shown in Step 2 of \autoref{fig:pipeline}, TraceMind traverses the layout and visibility snapshots and retrieves every recorded layout in which the semantic occurrence appears. We refer to these occurrence-specific layout records as \textbf{layout observations}. For each observation, TraceMind locates the occurrence's text span in the recorded layout and recovers its rendered geometry. The observation also retains the corresponding Chat or Draft scope, interface surface and page instance, text content, and visibility condition.

TraceMind reconstructs occurrence states by grouping layout observations that provide equivalent spatial anchors. Observations are assigned to the same state only when they refer to the same semantic occurrence and preserve the same interface context, text content, visibility condition, and approximately the same bounding-box shape. As illustrated in \autoref{fig:occurrence-states}, layouts that differ only because scrolling moves the occurrence within the viewport are merged. In contrast, text reflow changes the bounding-box shape and therefore produces a new state because it alters the spatial relationship between the target text and nearby interaction traces. Changes in content, interface context, or visibility likewise produce separate states.

For each state with observable geometry, TraceMind retains a \textbf{hard bounding box} that tightly encloses the rendered target text. It also derives a \textbf{soft bounding box} by expanding the hard box by a radius $\rho$ on all sides. The hard box anchors the occurrence itself, whereas the soft box preserves nearby pointer activity that may reflect attention without entering the exact text region. These state-specific boxes support the spatial alignment described in Section~\ref{section: State-Specific Pointer-Trajectory Heatmaps}.

\subsection{Aligning and Encoding Interaction Traces}
\label{section: Aligning and Encoding Interaction Traces}
As shown in Step 3 of \autoref{fig:pipeline}, TraceMind relates the logged pointer and operation events to the target unit's semantic occurrences and reconstructed occurrence states. This produces two records: target-relative pointer trajectories for each occurrence state and a target-linked operation sequence spanning the session. TraceMind encodes these records into three complementary representations: state-specific pointer-trajectory heatmaps, a time-binned activity sequence, and a workflow-informed feature vector.

\subsubsection{State-Specific Pointer-Trajectory Heatmaps}
\label{section: State-Specific Pointer-Trajectory Heatmaps}

\begin{figure*}[t]
    \centering
    \includegraphics[width=0.6\textwidth]{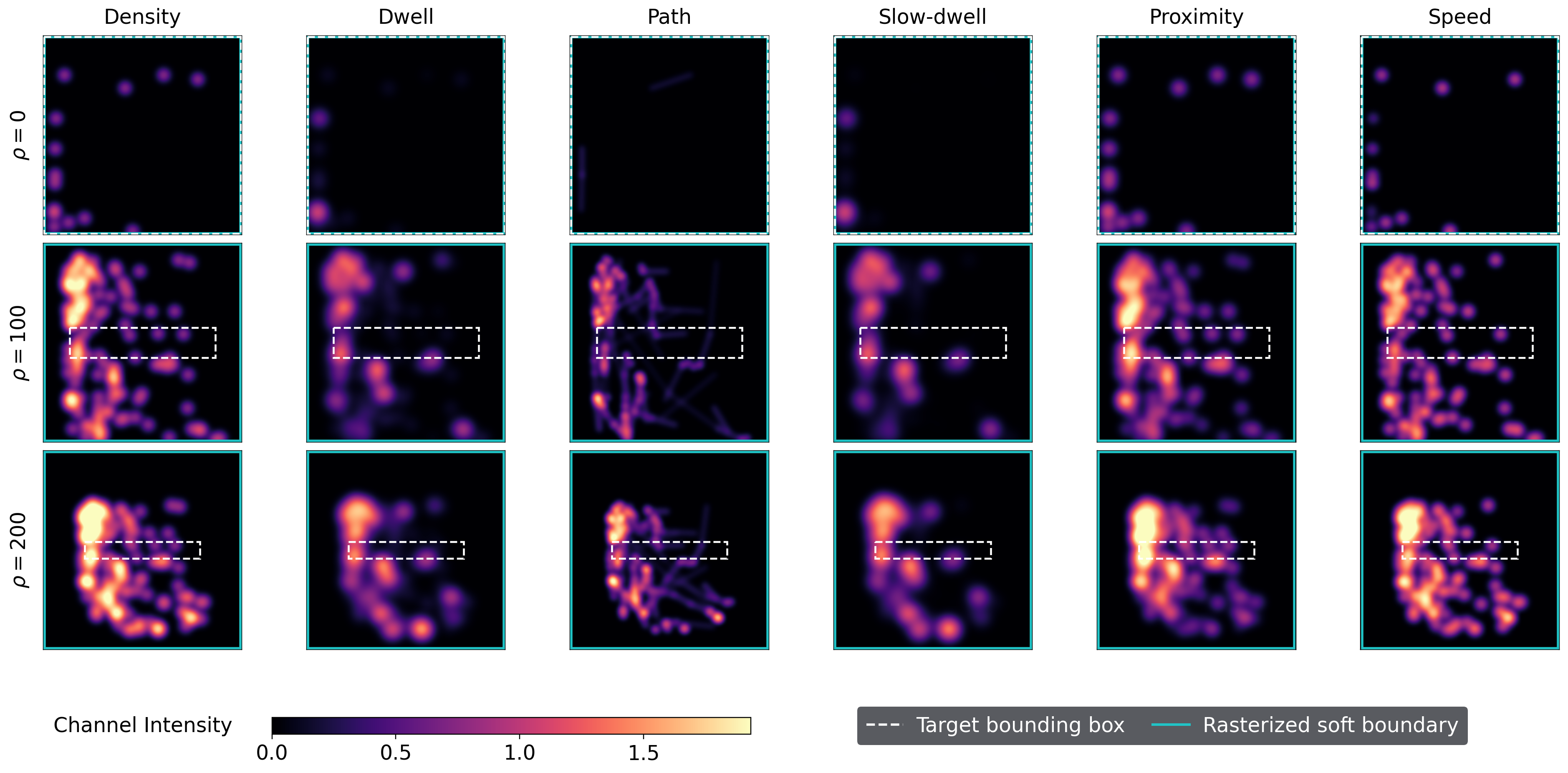}
    \caption{An example of spatial mouse-trajectory encoding for one occurrence state under three soft bounding box radii. Columns show the six heatmap channels; white dashed boxes mark the strict target box, cyan outlines mark the rasterized soft boundary.}
    \label{fig:six-channel-heatmaps}
\end{figure*}

Pointer positions are meaningful only relative to the layout in which they were recorded. TraceMind therefore aligns pointer coordinates to the target and encodes the resulting trajectories separately for each occurrence state. For state $s$, the hard bounding box $B_s=(l_s,t_s,r_s,b_s)$ provides the local spatial anchor, with width $w_s=r_s-l_s$ and height $h_s=b_s-t_s$.

Requiring the pointer to enter the hard bounding box would omit nearby movements that may still be associated with the target. We instead use the \textbf{soft bounding box} defined in Section~\ref{section: Reconstructing Layout-Aware Occurrence States for Each Semantic Occurrence} and retain pointer points inside the region obtained by expanding $B_s$ by $\rho$ on all sides:
\begin{equation}
\mathcal{P}_s(\rho)=\left\{p_i:\ l_s-\rho\leq x_i\leq r_s+\rho\ \text{and}\ t_s-\rho\leq y_i\leq b_s+\rho\right\}, \rho\geq0.
\label{eq:trajectory-soft-boundary}
\end{equation}
When $\rho=0$, the soft and hard bounding boxes coincide. Increasing $\rho$ retains a broader target-relative interaction context. We compare the effect of $\rho$ in Section~\ref{section: MODEL EVALUATION}.

To obtain a fixed-size spatial representation, we rasterize the retained trajectory as a $128\times128$ heatmap. Each retained pointer point $p_i=(x_i,y_i)$ is mapped to heatmap coordinates $\pi_{\rho}(p_i)=(\bar{x}_i,\bar{y}_i)$, where
\begin{equation}
\pi_{\rho}(p_i)=\left(
\frac{x_i-(l_s-\rho)}{w_s+2\rho}(W-1),
\frac{y_i-(t_s-\rho)}{h_s+2\rho}(H-1)
\right).
\label{eq:trajectory-grid-mapping}
\end{equation}
Here, $W=H=128$ denotes the spatial grid resolution. This transformation independently normalizes the horizontal and vertical extents of the expanded region to the heatmap frame.

Each heatmap contains six channels: density, dwell, path, slow dwell, target proximity, and speed. For the five point-based channels, we distribute each point's channel-specific weight over nearby pixels with a Gaussian kernel. For the path channel, we rasterize the line between consecutive points. Let $\mathcal{E}_s(\rho)$ contain consecutive point pairs from the same trajectory segment that are separated by at most three seconds. The value of channel $c$ at heatmap location $q$ is
\begin{equation}
\widetilde{H}_{s,c}(q)=
\begin{cases}
\displaystyle\sum_{p_i\in\mathcal{P}_s(\rho)}\omega_{i,c}G_{\sigma_c}\!\left(q-\pi_{\rho}(p_i)\right), & c\neq\mathrm{path},\\[6pt]
\displaystyle\sum_{(p_i,p_j)\in\mathcal{E}_s(\rho)}L_{\sigma_c}\!\left(q;\pi_{\rho}(p_i),\pi_{\rho}(p_j)\right), & c=\mathrm{path},
\end{cases}
\label{eq:trajectory-channel-accumulation}
\end{equation}
where $G_{\sigma_c}$ is a Gaussian kernel centered at the mapped point and weighted by $\omega_{i,c}$, and $L_{\sigma_c}$ rasterizes the line between two consecutive mapped points.

Because repeated activity can produce highly skewed channel values, we further apply
\begin{equation}
H_{s,c}(q)=\log\!\left(1+\widetilde{H}_{s,c}(q)\right),
\label{eq:trajectory-log-scaling}
\end{equation}
which compresses large intensities while preserving zero-valued regions. The resulting $H_s\in\mathbb{R}^{6\times128\times128}$ represents the target-relative pointer trajectory for state $s$. Alongside each heatmap, TraceMind retains a 16-dimensional state summary covering interaction amount, target proximity, relative position, interface scope, etc. For each information unit, we retain up to 16 occurrence states and use a binary mask to distinguish observed states from padding.

\autoref{fig:six-channel-heatmaps} illustrates how increasing the soft bounding box radius $\rho$ expands the retained spatial context across the six heatmap channels.

\subsubsection{Time-Binned Activity Sequence}
The interaction log contains both frequent pointer events and sparse operations such as selection, copying, pasting, and editing. To preserve their order without allowing high-frequency events to dominate, TraceMind represents the target-linked operations on a normalized session timeline.

TraceMind first constructs event–target relations. Text operations (e.g., select, copy, paste, text insertion) are linked to the target unit when their affected range uniquely overlaps one of its semantic occurrences. Pointer-based events (e.g., pointer move, pointer click) are linked through the available occurrence-state geometry at the event time to measure the continuous distance from the pointer to the rendered target text. It converts this distance into proximity signals at multiple spatial scales, assigning stronger target-related evidence to closer pointer events. 

We divide each session into $T=512$ equal-width bins and construct, for each information unit, a time-binned activity sequence $X^{\mathrm{op}}\in\mathbb{R}^{512\times44}$. The vector for bin $k$ is
\begin{equation}
\mathbf{x}^{\mathrm{op}}_k=
[\mathbf{s}_k;\mathbf{r}_k;\mathbf{a}_k].
\end{equation}
Here, $\mathbf{s}_k\in\mathbb{R}^{20}$ summarizes the interaction context within the bin, including pointer movements, clicks, scrolling, editing, selection, clipboard operations, visibility changes, and surface transitions. $\mathbf{r}_k\in\mathbb{R}^{16}$ summarizes the constructed event-target relations, including text- and pointer-relation counts, Chat or Draft scope, source context, layout observability, and distance-based proximity. Finally, $\mathbf{a}_k\in\mathbb{R}^{8}$ captures bin-level conjunctions between activity types and target-related evidence, such as copying, editing, clicking, or scrolling in a time bin where such evidence is present. We apply $\log(1+x)$ to count-like values and use a binary mask to distinguish observed from empty bins.

\subsubsection{Workflow-Informed Interaction Features}
\label{section: Workflow-Informed Interaction Features}
The time-binned sequence preserves local temporal progression, but does not directly represent sparse, high-level behavioral patterns that span distant events, such as revisiting a source after editing the Draft, or copy-and-paste-based text transfer. TraceMind therefore uses the same target-linked operation sequence to derive a workflow-informed feature vector that summarizes higher-level engagement with the target unit.

We developed these features through iterative inspection of 15 pilot co-generation session replays collected before the main data collection study. These sessions were not included in the 62-session dataset described in Section~\ref{section: Dataset Construction}. Two HCI researchers independently reviewed how users interacted with individual information units across Chat and Draft, without access to post-task recognition outcomes. The researchers captured recurrent patterns of observable user engagement. We retained recurrent patterns that could be computed objectively from the interaction logs.

The resulting 40-dimensional feature set captures eight behavior families: Chat-side repeated inspection, Draft-side repeated inspection, Draft-side inspection after editing, Draft editing and revision, copy-and-paste-based transfer, cross-surface text selection, cross-surface coordination, and source checking before revision. These behavior families also broadly align with prior work that characterizes rereading, revisitation, and content manipulation as observable traces of information processing \cite{rakovic2023harnessing, leroy2023investigating}. The complete feature set is provided in Appendix~\ref{appendix: Workflow-informed Interaction Features}. For each information unit, these features form a vector $\mathbf{x}^{\mathrm{wf}}\in\mathbb{R}^{40}$. Count- and duration-based features are transformed using $\log(1+x)$ before modeling.

\subsection{Information-Uptake Prediction Model}
\label{section: Information-Uptake Prediction Model}
The three representations above have different structures and capture complementary interaction evidence. As shown in Step 4 of \autoref{fig:pipeline}, TraceMind models each representation in a separate branch and combines the three branch probabilities through weighted-logit late fusion. This maps the heterogeneous signals into a common prediction space, enabling simple and effective fusion.

\textbf{Spatial trajectory branch.} TraceMind first applies channel-wise mean and maximum pooling across the valid state-specific trajectory heatmaps. Their concatenation forms a 12-channel image that preserves both recurring and salient spatial patterns. A residual CNN with squeeze-and-excitation gating encodes this image into a 128-dimensional representation. In parallel, TraceMind averages the 16-dimensional state summaries across valid states and projects the resulting vector to 48 dimensions. The two representations are concatenated and passed through a linear classification head to produce the spatial-branch uptake probability $p_{\mathrm{traj}}$. We train this branch with class-weighted binary cross-entropy and use light channel dropout and Gaussian perturbation for regularization.

\textbf{Temporal activity branch.} TraceMind first projects each bin of $X^{\mathrm{op}}$ into a 48-dimensional hidden space. A three-layer temporal convolutional network (TCN), with dilation rates 1, 2, and 4, models interaction patterns over multiple temporal scales. Mask-aware attentive pooling then aggregates the observed bins into a temporal embedding. We first pre-train the TCN encoder through fold-local masked-bin reconstruction: 18\% of the observed bins are masked, and a temporary decoder reconstructs their original interaction features from the surrounding temporal context. After pre-training, we discard the decoder and jointly fine-tune the encoder, attentive-pooling layer, and linear classification head using class-weighted binary cross-entropy. At inference time, the trained model processes the complete unmasked sequence and outputs the temporal-branch uptake probability $p_{\mathrm{temp}}$.

\textbf{Workflow-informed feature branch.} TraceMind models the 40-dimensional workflow-informed feature vector using ElasticNet-regularized logistic regression. Count- and duration-based features are log-transformed, missing values are imputed with training-set medians, and all features are standardized using training-set statistics. The model outputs the workflow-branch uptake probability $p_{\mathrm{wf}}$.

TraceMind combines the three probabilities by taking a positive weighted sum of their log-odds:
\begin{equation}
\hat{p}=\sigma\left(
\alpha_{\mathrm{traj}}\operatorname{logit}(p_{\mathrm{traj}})
+\alpha_{\mathrm{temp}}\operatorname{logit}(p_{\mathrm{temp}})
+\alpha_{\mathrm{wf}}\operatorname{logit}(p_{\mathrm{wf}})
\right),
\label{eq:weighted-logit-fusion}
\end{equation}
where $\sigma$ is the sigmoid function, $\alpha_{\mathrm{traj}},\alpha_{\mathrm{temp}},\alpha_{\mathrm{wf}}>0$, and $\alpha_{\mathrm{traj}}+\alpha_{\mathrm{temp}}+\alpha_{\mathrm{wf}}=1$. TraceMind predicts uptake when $\hat{p}\geq t$ and non-uptake otherwise. The decision threshold $t$ is selected using only the training data and fixed before evaluation on the held-out test set.

%% file: data/chap5.tex
\section{Model Evaluation}
\label{section: MODEL EVALUATION}
In this section, we evaluate TraceMind using the dataset collected in Section~\ref{section: DATA COLLECTION STUDY}. We next describe the evaluation setup, overall performance, and ablation results.

\subsection{Experimental Setup}
\subsubsection{Evaluation Protocol}
We evaluate TraceMind on the dataset collected in Section~\ref{section: DATA COLLECTION STUDY}, containing 1187 labeled information units from 62 participants. We use 5-fold grouped cross-validation with participant ID as the grouping variable, ensuring that all units from the same participant remain in the same split and avoiding participant-level leakage. We aggregate predictions from the held-out folds for final evaluation.

\subsubsection{Baseline Models}
We compared TraceMind with the following baseline models.

\begin{itemize}
    \item \textbf{Content \& Context + ElasticNet}: Prior work shows that content and task context can affect interaction-based judgments independently of observed behavior \cite{mao2017understanding}. This baseline combines a frozen embedding of the information-unit text with non-behavioral attributes, including text length, normalized position in the final draft, provenance, and task type. The resulting features are classified using ElasticNet-regularized logistic regression.
    \item \textbf{Conventional Interaction + GBT}: Conventional interaction statistics, such as exposure time, scrolling, and pointer activity, have been used to predict reading difficulty and content usefulness \cite{gooding2021predicting,mao2017understanding,zhu2024modeling}. This baseline summarizes each information unit using its viewport exposure, pointer dwell and movement, clicks, scrolling, selection, copy, paste, and editing counts, and trains a Gradient Boosting Tree (GBT) classifier. It excludes the revisit, cross-surface, and source-checking relations of our workflow-informed feature set.
    \item \textbf{Workflow-Informed Features + GBT}: This baseline uses the 40-dimensional workflow-informed feature vector $\mathbf{x}^{\mathrm{wf}}$ in Table~\ref{tab:workflow-informed-features} as input to GBT.
    \item \textbf{LSTM + DNN}: Recurrent models have been widely used to encode mouse interaction sequences \cite{arapakis2020learning}. This baseline processes the time-binned interaction representation $X^{\mathrm{op}}\in\mathbb{R}^{512\times44}$ using three Long Short-Term Memory (LSTM) layers with 128, 64, and 32 hidden units. The resulting temporal representation is concatenated with workflow-informed features and classified by a Deep Neural Network (DNN).
    \item \textbf{Transformer + DNN}: Transformer encoders have also been used to model cursor trajectories for attention prediction \cite{villaizan2025adsight}. This baseline applies a Transformer encoder with sinusoidal positional encodings to $X^{\mathrm{op}}$ and reads out the temporal representation from a [CLS] token. The Transformer is trained from scratch, without the masked-bin pre-training used by TraceMind's temporal branch. The other components are consistent with the baseline of \textbf{LSTM + DNN}. 
    \item \textbf{ResNet-50 + DNN}: Image-based encodings paired with ResNet-50 provide a strong visual baseline for mouse-movement modeling \cite{arapakis2020learning,villaizan2025adsight}. This baseline adapts the first convolutional layer of an ImageNet-pretrained ResNet-50 to the six-channel trajectory heatmaps described in Section~\ref{section: State-Specific Pointer-Trajectory Heatmaps}, encodes each valid occurrence state, and mean-pools the state representations. The other components are consistent with the baseline of \textbf{LSTM + DNN}.
    \item \textbf{Mouse2Vec + DNN}: Mouse2Vec learns reusable mouse-behavior representations through self-supervised pretraining \cite{zhang2024mouse2vec}. This baseline applies the frozen Mouse2Vec encoder to the raw pointer segments associated with an information unit and pools the resulting segment embeddings. The other components are consistent with the baseline of \textbf{LSTM + DNN}.
    \item \textbf{Full-Input Early-Fusion DNN}: This baseline receives the same spatial trajectory, temporal interaction, and workflow-informed inputs as TraceMind. It concatenates the spatial and temporal embeddings with the workflow-informed feature vector and predicts uptake using a DNN. The temporal encoder uses the same masked-bin pretraining as TraceMind.
\end{itemize}

\begin{table*}[b]
    \centering
    \caption{Overall performance of TraceMind and the baselines. All results are presented as percentages (\%). Bold indicates TraceMind, and underlining marks the strongest learned baseline for each metric.}
    \label{tab:overall-performance}
    \setlength{\tabcolsep}{7pt}
    \renewcommand{\arraystretch}{1.08}
    \small
    \begin{tabular}{lcccc}
        \toprule
        \textbf{Model} &
        \textbf{AUROC} &
        \textbf{Bal. Acc.} &
        \textbf{AUPRC$_{\mathrm{non}}$} &
        \textbf{Macro-F1} \\
        \midrule
        Content \& Context + ElasticNet & 65.01 & 60.56 & 27.12 & 53.00 \\
        Conventional Interaction + GBT & 62.33 & 61.75 & 32.05 & 60.25 \\
        Workflow-Informed Features + GBT & 71.42 & 62.57 & 33.96 & 45.14 \\
        LSTM + DNN & 75.58 & 65.84 & 44.01 & \underline{64.84} \\
        Transformer + DNN & 65.81 & 63.46 & 28.28 & 53.05 \\
        ResNet-50 + DNN & 75.00 & 60.09 & 43.17 & 61.73 \\
        Mouse2Vec + DNN & 72.58 & \underline{68.20} & 37.20 & 62.76 \\
        Full-Input Early-Fusion DNN & \underline{76.68} & 58.62 & \underline{45.36} & 59.88 \\
        \midrule
        \textbf{TraceMind} & \textbf{81.17} & \textbf{80.58} & \textbf{49.34} & \textbf{73.05} \\
        \bottomrule
    \end{tabular}
    \vspace{2pt}
\end{table*}

\subsubsection{Evaluation Metrics}
Because our dataset contains significantly more uptake labels than non-uptake labels, conventional accuracy can overemphasize the majority class. We therefore report balanced accuracy and macro-F1 for thresholded predictions, which give equal importance to the two classes while capturing class-wise recall and precision \cite{brodersen2010balanced, sokolova2009systematic}. We also report the area under the precision-recall curve for non-uptake $\mathrm{AUPRC}_{\text{non}}$. This metric evaluates the precision-recall tradeoff for detecting the less frequent non-uptake cases, which are particularly important for identifying information that may require additional review \cite{davis2006relationship, saito2015precision}. We also report AUROC to measure the model's overall ability to distinguish uptake from non-uptake across decision thresholds.

\subsection{Overall Performance}
As shown in \autoref{tab:overall-performance}, TraceMind achieved the strongest overall performance among the learned models across all four metrics. Compared with the strongest learned baseline for each metric, TraceMind improved AUROC by 4.49 percentage points, balanced accuracy by 12.38 points, $\mathrm{AUPRC}_{\mathrm{non}}$ by 3.98 points, and macro-F1 by 8.21 points.

\begin{figure}[]
    \centering
    \includegraphics[width=0.5\textwidth]{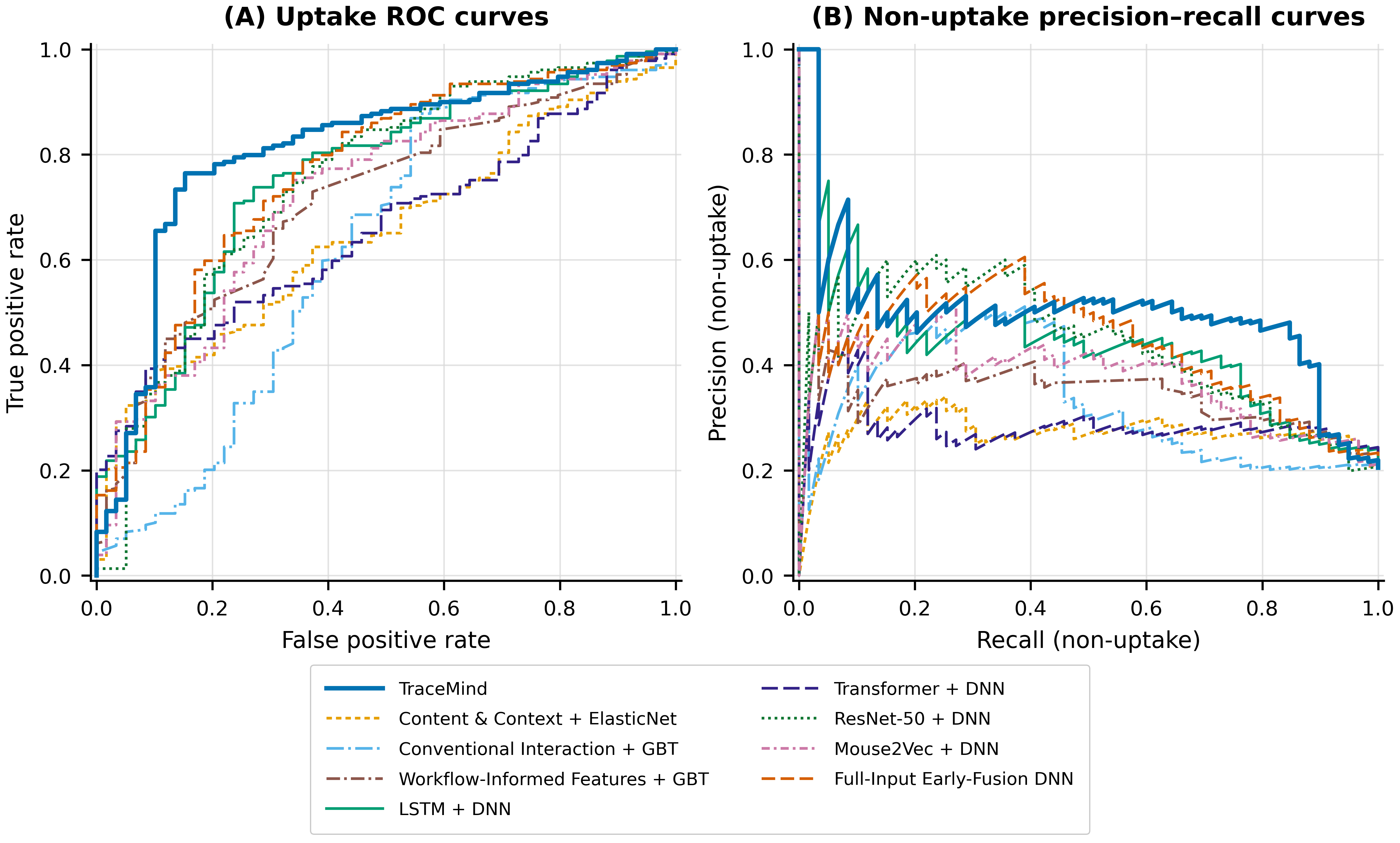}
    \caption{Performance of TraceMind and the baselines. (A) ROC curves treating uptake as the positive class. (B) Precision-recall curves treating the less frequent non-uptake class as positive.}
    \label{fig:overall-performance-curves}
\end{figure}

The strongest competing baseline varied across metrics: Full-Input Early-Fusion DNN performed best on AUROC and $\mathrm{AUPRC}_{\mathrm{non}}$, Mouse2Vec + DNN performed best on balanced accuracy, and LSTM + DNN performed best on macro-F1. As shown in \autoref{fig:overall-performance-curves}, TraceMind's advantage was also sustained across decision thresholds. 

\subsection{Ablation Results}
\label{section: Ablation Results}
We next conduct a series of ablation analyses on TraceMind to examine the contributions of its three input branches and other key design choices.

\begin{table*}[htbp]
    \centering
    \caption{Ablation results across TraceMind's input branches, interaction traces, alignment mechanisms, and TCN pretraining for the temporal branch. All results are percentages (\%). }
    \label{tab:combined-ablation}
    \setlength{\tabcolsep}{4.5pt}
    \renewcommand{\arraystretch}{1.05}
    \small
    \begin{tabular}{llcccc}
        \toprule
        \textbf{Ablated Design} &
        \textbf{Model Variant} &
        \textbf{AUROC} &
        \textbf{Bal. Acc.} &
        \textbf{AUPRC$_{\mathrm{non}}$} &
        \textbf{Macro-F1} \\
        \midrule
        \multirow{7}{*}{Input branches}
        & Spatial only  & 77.91 & 69.69 & 43.96 & 58.61 \\
        & Temporal only & 78.00 & 72.85 & 43.77 & 65.41 \\
        & Workflow only & 74.69 & 72.14 & 40.32 & 63.18 \\
        & w/o Spatial   & 80.45 & 77.36 & 44.85 & 69.75 \\
        & w/o Temporal  & 79.11 & 71.54 & 47.61 & 63.64 \\
        & w/o Workflow  & 80.25 & 73.95 & 48.76 & 61.62 \\
        \midrule
        \multirow{2}{*}{Interaction traces}
        & Chat-only traces  & 75.49 & 67.94 & 39.84 & 56.30 \\
        & Draft-only traces & 75.96 & 73.87 & 40.36 & 64.86 \\
        \midrule
        \multirow{2}{*}{Alignment}
        & w/o semantic alignment     & 72.91 & 70.36 & 33.21 & 65.09 \\
        & w/o layout-state alignment & 80.70 & 78.01 & 45.25 & 70.70 \\
        \midrule
        \multirow{2}{*}{TCN pretraining}
        & Temporal only (w/o TCN pretraining) & 76.07 & 69.76 & 36.01 & 65.43 \\
        & TraceMind (w/o TCN pretraining)       & 79.94 & 72.27 & 43.60 & 66.51 \\
        \midrule
        \multicolumn{1}{c}{\multirow{1}{*}{---}}
        & \textbf{TraceMind} & \textbf{81.17} & \textbf{80.58} & \textbf{49.34} & \textbf{73.05} \\
        \bottomrule
    \end{tabular}
\end{table*}

\subsubsection{Contribution of Three Input Branches}
To assess the contribution of each input branch, we evaluate the performance of each branch individually and the overall model performance when each branch is removed in turn.

As shown in \autoref{tab:combined-ablation} (Input branches block), all three branches provided predictive signal independently, with AUROC ranging from 74.69\% to 78.00\%. Moreover, removing the temporal branch produced the largest decreases in AUROC and balanced accuracy (2.06 and 9.04 percentage points), removing the workflow-informed branch produced the largest decrease in macro-F1 (11.43 points), and removing the spatial branch produced the largest decrease in $\mathrm{AUPRC}_{\mathrm{non}}$ (4.49 points). These results suggest that the three branches capture distinct and complementary aspects of information uptake.

\subsubsection{Contribution of Chat and Draft Interaction Traces}
\label{section: Contribution of Chat and Draft Interaction Traces}
To examine how interactions across the two interfaces contribute to prediction, we restricted all three branches to either Chat-only or Draft-only traces, and omitted cross-surface workflow features. All model components and training settings were kept unchanged.

As shown in \autoref{tab:combined-ablation} (Interaction traces block), Draft-only traces retained more predictive signal than Chat-only traces. Moreover, combining interaction traces from both interfaces consistently outperformed using either interface alone. Compared with using Chat-only and Draft-only traces, TraceMind improved AUROC by 5.68 and 5.21 points, balanced accuracy by 12.64 and 6.71 points, $\mathrm{AUPRC}_{\mathrm{non}}$ by 9.50 and 8.98 points, and macro-F1 by 16.75 and 8.19 points, respectively. These results suggest that Draft interactions provide the stronger individual signal, while Chat interactions contribute complementary evidence to the full model.

\subsubsection{Contribution of Semantic and Layout-state Alignment}
\label{section: Contribution of Semantic and Layout-state Alignment}
To examine the contribution of the two alignment stages, we evaluated two ablated versions of TraceMind. For the semantic-alignment ablation, we rebuilt each information unit’s unit-linked inputs solely from its final-draft occurrence, excluding matched occurrences in Chat and earlier Draft versions, while retaining the temporal branch’s session-level context. For the layout-state ablation, we retained all identified semantic occurrences and the bbox-anchored trajectory encoding, but merged the layout states belonging to each occurrence into one occurrence-level trajectory image. All other model components and training settings were kept unchanged.

As shown in \autoref{tab:combined-ablation} (Alignment block), using only the final-draft anchor reduced AUROC by 8.26 percentage points, balanced accuracy by 10.22 points, $\mathrm{AUPRC}_{\mathrm{non}}$ by 16.13 points, and macro-F1 by 7.96 points. Removing layout-state alignment had a smaller effect after fusion, but still reduced balanced accuracy by 2.57 points, $\mathrm{AUPRC}_{\mathrm{non}}$ by 4.09 points, and macro-F1 by 2.35 points. Its effect was more pronounced within the spatial trajectory branch, whose AUROC decreased from 77.91\% to 74.12\%. Together, these results suggest that recovering an information unit's semantic occurrences across Chat and Draft provides substantial predictive signal beyond its final-draft occurrence alone, while layout-state alignment preserves complementary spatial evidence that the other branches can partially compensate for.

\subsubsection{Contribution of TCN Pretraining on Temporal Patterns Branch} 
To isolate the contribution of TCN pretraining, we removed the masked-bin-reconstruction pretraining stage and trained the temporal TCN from scratch. All other model components and training settings were kept unchanged.

As shown in \autoref{tab:combined-ablation} (TCN pretraining block), pretraining improved the temporal branch's AUROC by 1.93 percentage points and $\mathrm{AUPRC}_{\mathrm{non}}$ by 7.76 points. Its benefit persisted after fusion: compared with training the TCN from scratch, the full TraceMind model improved AUROC by 1.23 points, balanced accuracy by 8.31 points, $\mathrm{AUPRC}_{\mathrm{non}}$ by 5.74 points, and macro-F1 by 6.54 points. These results suggest that masked-bin-reconstruction pretraining helps the TCN learn temporal representations that complement the other two branches.

\subsubsection{Effect of Soft Boundary} 
To examine how much spatial context should be retained around each rendered occurrence, we varied the bbox-anchored soft-boundary radius $\rho$ and rebuilt the spatial trajectory input for each setting. All other model components and training settings were kept unchanged.

\begin{figure}[htbp]
    \centering
    \includegraphics[width=1.0\linewidth]{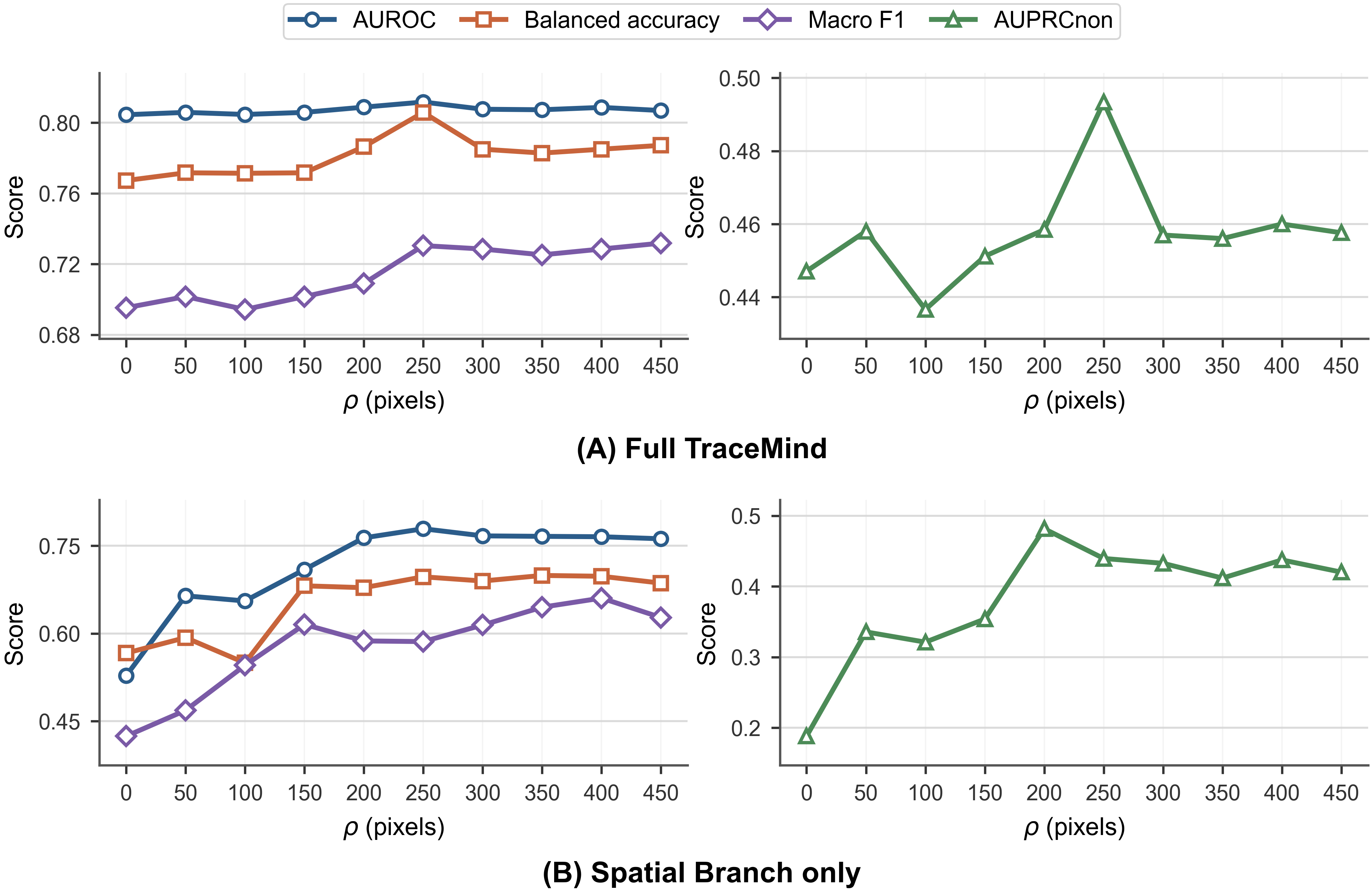}
    \caption{The effect of soft boundary radius.}
    \label{fig:soft-boundary-analysis}
\end{figure}

As shown in \autoref{fig:soft-boundary-analysis}, restricting trajectories to the target bounding box ($\rho=0$) substantially weakened the spatial branch, whose AUROC decreased to 52.77\%. Performance improved as nearby movements were included, with $\rho=250$ yielding the strongest overall fusion results. Expanding the boundary further did not consistently improve the fused model. These results suggest that pointer movements near an information unit provide useful evidence beyond direct entry into its bounding box, while overly broad regions add limited additional signal.

%% file: data/chap6.tex
\section{Behavioral Analysis}
Beyond evaluating TraceMind's predictive performance, we further conduct a behavioral analysis on the collected data to examine user behavioral patterns associated with information uptake during human–LLM co-generation.

\begin{figure*}[]
    \centering
    \includegraphics[width=1.0\linewidth]{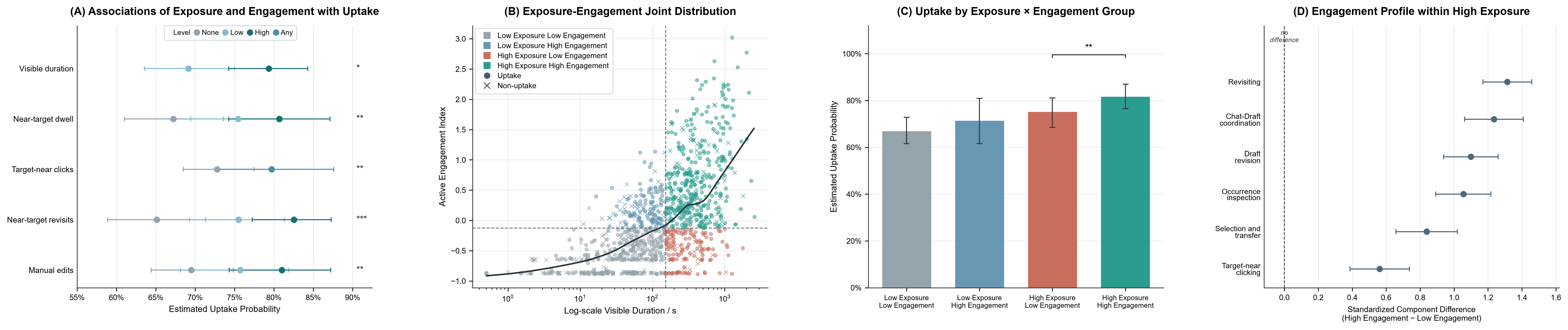}
    \caption{Relationships among information exposure, active engagement, and information uptake at both the individual-signal and joint-pattern levels. (A) Estimated uptake probabilities across predefined levels of visible duration and representative active-engagement signals. (B) Joint distribution of visible duration and active engagement across information units; dashed lines indicate the median splits, while circles and crosses denote uptake and non-uptake, respectively. (C) Estimated uptake probabilities for the four exposure--engagement groups. (D) Standardized differences in engagement components between High and Low Engagement units within the high-exposure group; positive values indicate higher levels among High Engagement units.}
    \label{fig:exposure-engagement-analysis}
\end{figure*}

\subsection{Information Uptake Across Exposure Duration and Active Engagement}
\label{section: Information Uptake Across Exposure Duration and Active Engagement}
In this section, we examine how exposure duration and active engagement were associated with information uptake. We first examined whether greater exposure duration and individual active-engagement signals were associated with higher information uptake. For signals with sufficiently many positive observations (e.g., visible duration and target-near dwell), we divided units into low and high groups at the median. We then estimated task- and sample-kind-adjusted uptake probabilities using logistic standardization, with 95\% CIs obtained from 1,000 participant-cluster bootstrap samples. As shown in \autoref{fig:exposure-engagement-analysis}A, uptake was significantly higher in the high-exposure group than in the low-exposure group ($p<0.05$). Higher active engagement was also associated with greater uptake, with significant differences for target-near dwell ($p<0.01$), clicks ($p<0.01$), revisits ($p<0.001$), and manual edits ($p<0.01$). Overall, both greater exposure and more active engagement were positively associated with information uptake, with stronger evidence for the active-engagement signals.

We then examined the joint pattern of exposure duration and active engagement. We constructed an equal-weight active-engagement index from six standardized components: target-near clicking, revisiting, occurrence inspection, Chat--Draft coordination, Draft revision, and selection and transfer. We divided all units at the global medians of visible duration and the engagement index, retaining every unit in one of four groups: Low Exposure Low Engagement, Low Exposure High Engagement, High Exposure Low Engagement, and High Exposure High Engagement. As shown in \autoref{fig:exposure-engagement-analysis}B, visible duration and the engagement index showed a moderate overall positive correlation ($r_s=.61$), with the concordant groups (High Exposure High Engagement and Low Exposure Low Engagement) comprising 72.2\% of units and the discordant groups (High Exposure Low Engagement and Low Exposure High Engagement) comprising 27.8\%.

We further compared information uptake across the four groups. We estimated uptake probabilities using task-adjusted logistic standardization, with 95\% CIs obtained from 1,000 participant-cluster bootstrap samples. We then compared High and Low Engagement within each exposure level (\autoref{fig:exposure-engagement-analysis}C). High Engagement showed higher estimated uptake than Low Engagement under both low exposure duration (71.5\% vs. 67.0\%) and high exposure duration (81.8\% vs. 75.3\%). Participant-clustered GEE tests with Holm correction showed that this increase was significant within the high-exposure group ($p<0.01$), but not within the low-exposure group.

Within the high-exposure group, we further examined which engagement behaviors distinguished High and Low Engagement units, whose uptake probabilities differed significantly (\autoref{fig:exposure-engagement-analysis}D). The largest adjusted standardized differences were observed in revisiting ($1.31SD$), Chat--Draft coordination ($1.24SD$), Draft revision ($1.10SD$), and occurrence inspection ($1.05SD$).

Together, exposure duration and active engagement were each positively associated with information uptake and were also positively associated with each other. Estimated uptake was highest when both were high. Among highly exposed units, more sustained and coordinated engagement was associated with higher uptake.

\begin{figure*}[]
    \centering
    \includegraphics[width=1.0\linewidth]{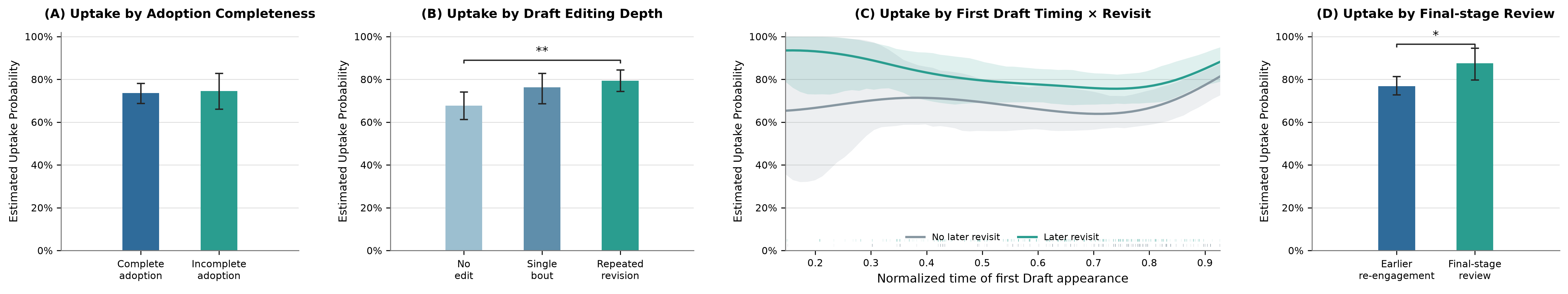}
    \caption{Information uptake across adoption completeness and subsequent engagement. (A) Uptake probabilities for complete and incomplete Chat adoption. (B) Uptake probabilities by Draft editing depth among the Chat-adopted units. (C) Covariate-adjusted uptake over normalized time of first Draft appearance, separated by whether the unit was later revisited in the Draft. (D) Covariate-adjusted uptake for earlier re-engagement and final-stage review among re-engaged units.}
    \label{fig:adoption-uptake-analysis}
\end{figure*}

\subsection{Information Uptake Across Content Adoption Completeness and Post-Adoption Engagement}
\label{section: Information Uptake Across Content Adoption Completeness and Post-Adoption Engagement}
In this section, we examine whether information uptake varied with how LLM-provided content entered the Draft and how users subsequently engaged with it.
We first categorized each valid information unit based on whether and how it was adopted from Chat into the Draft. For each unit, we used an LLM judge (GPT-5.4-mini) to determine whether its first Draft occurrence was semantically matched to an earlier assistant message. Units with no such match were classified as \textbf{non-adoption}. For matched units, we further distinguished \textbf{complete adoption}, where the normalized Draft occurrence appeared exactly as written in the matched Chat message, from \textbf{incomplete adoption}, where the same information was transferred with rephrasing. Given the small number of non-adoption units (13 units from nine participants, all showing uptake), our subsequent analysis only focuses on the Chat-adopted units, including complete and incomplete adoption units.

To begin with, complete and incomplete adoption showed nearly identical uptake probabilities (\autoref{fig:adoption-uptake-analysis}A): 73.7\% for complete adoption and 74.5\% for incomplete adoption. Thus, preserving the assistant's exact wording alone was not significantly associated with uptake.

Post-adoption Draft engagement showed a clearer association with uptake (\autoref{fig:adoption-uptake-analysis}B). The overall effect of editing depth was significant ($p<0.01$). Among Chat-adopted units, uptake increased from 67.7\% with no target edit to 76.3\% after a single edit bout and 79.4\% after repeated revision. 

For the timing analyses, we estimated standardized logistic predictions over the normalized time of first Draft appearance (\autoref{fig:adoption-uptake-analysis}C). The timing of initial insertion did not show a reliable association with uptake. By contrast, the timing of subsequent engagement showed a clearer association (\autoref{fig:adoption-uptake-analysis}D). Among units with later re-engagement, we classified a unit as receiving a final-stage review when its last target-related interaction occurred within the final 10\% of the task or within 120 seconds before final draft submission; all other cases were classified as earlier re-engagement. Units receiving a final-stage review had a higher uptake probability than those whose re-engagement occurred earlier ($p<0.05$).

Together, these results suggest that uptake was associated less with how or when information entered the Draft than with how it was subsequently processed. Specifically, exact wording and initial timing did not reliably distinguish uptake, whereas repeated revision and final-stage review were associated with higher uptake.

\subsection{Self-reported Confidence and Metacognitive Awareness of Information Uptake}
\label{section: Self-reported Confidence and Metacognitive Awareness of Information Uptake}
In this section, we examine whether participants' self-reported confidence reflected their information uptake and whether interaction traces could reveal failures that confidence did not. Participants rated their confidence on a five-point scale after each recognition question. Here, we defined ratings of 4-5 as high confidence and an incorrect high-confidence response as a high-confidence miss. Confidence analyses used all 1187 valid units from 62 participants.

\begin{figure*}[]
    \centering
    \includegraphics[width=1.0\linewidth]{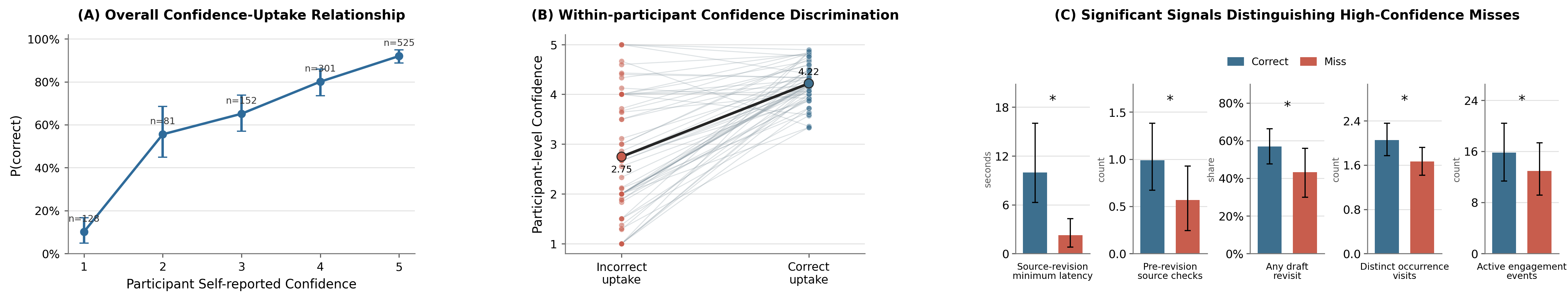}
    \caption{Relationship between participants’ self-reported confidence and information uptake, with behavioral signals further distinguishing high-confidence misses from correct uptake. (A) Unit-level uptake correctness by confidence rating; error bars indicate participant-cluster bootstrap 95\% CIs. (B) Participant-level mean confidence for incorrect and correct uptake units, with each line representing one participant. (C) Behavioral signals that significantly distinguished high-confidence misses from high-confidence correct units; bars show participant-level means with 95\% bootstrap CIs.}
    \label{fig:figure_confidence_and_within_person_sensitivity}
\end{figure*}

We found that self-reported confidence strongly reflected information uptake. Recognition accuracy increased monotonically from 10.2\% at confidence~1 to 92.0\% at confidence~5 (see \autoref{fig:figure_confidence_and_within_person_sensitivity}A). The same pattern held within participants: among the 58 participants with both correct and incorrect units, mean confidence was 4.22 for correct units versus 2.75 for incorrect units, with 50 participants showing higher confidence when correct ($p<0.001$; see \autoref{fig:figure_confidence_and_within_person_sensitivity}B). Thus, participants were generally sensitive to whether they had successfully taken up an information unit.

However, self-reported confidence did not capture all uptake failures. We observed 102 high-confidence misses, accounting for 12.3\% of high-confidence units and 33.3\% of all incorrect units. Moreover, 36 of 62 participants (58.1\%) experienced at least one high-confidence miss.

We further examined whether interaction traces could distinguish these failures from high-confidence correct units. Across the 40 workflow-informed interaction features (Section~\ref{section: Workflow-Informed Interaction Features}), five showed significant differences in participant-level paired comparisons ($p<0.05$), all indicating weaker engagement for high-confidence misses than for high-confidence correct units (see \autoref{fig:figure_confidence_and_within_person_sensitivity}C). Specifically, high-confidence misses showed fewer source checks before Draft revision ($M=0.57$ vs. $M=0.99$), a lower minimum interval from source checking to Draft revision ($M=2.25$~s vs. $M=9.98$~s), a lower likelihood of revisiting the unit in the Draft ($M=43.3\%$ vs. $M=56.9\%$), fewer distinct occurrences of the unit visited across Chat and Draft ($M=1.66$ vs. $M=2.05$), and fewer unit-linked engagement events (e.g., visits, copying, pasting, selection, editing operations) across the Chat and Draft surfaces ($M=12.92$ vs. $M=15.78$). Interaction traces may therefore provide complementary evidence of uptake risk when self-reported confidence alone is misleading.

%% file: data/chap7.tex
\section{Discussion}
In this section, we discuss the broader implications of our findings for understanding and modeling information uptake in human-LLM co-generation. First, we discuss information uptake as an unfolding interaction process, highlighting sustained active engagement as particularly informative for future HCI user modeling. Second, we discuss how performance-based uptake assessment can be extended to open-ended artifacts. Third, we discuss how low-cost interaction traces can support user-state modeling in dynamic everyday workflows. Fourth, we consider practical applications of our work for motivating the design of uptake-aware human-LLM co-generation systems. Finally, we discuss the limitations of this work and directions for future research.

\subsection{Information Uptake Unfolds Across the Interaction Process}
Across our modeling and behavioral analyses, evidence of information uptake was distributed across the evolving interaction history of an information unit rather than concentrated in a single event, interface, or moment. In our modeling results, preserving multiple semantic occurrences and layout states improved prediction, while combining Chat and Draft traces outperformed either surface alone (Section~\ref{section: Ablation Results}). Our behavioral analyses showed a similar pattern: exposure, content adoption, and self-reported confidence each provided incomplete evidence of uptake. Together, these results suggest that information uptake is an unfolding process reflected in how users continue to engage with the same information over time.

This process-level view can inform how HCI systems model user states from interaction behavior. Prior work often summarizes behavior into fixed representations, such as session-level interaction features for predicting knowledge gain \cite{yu2018predicting}, page-level statistics for modeling engagement \cite{grinberg2018identifying}, or task-instance signals for modeling AI reliance \cite{ma2023should}. Our findings suggest that these signals may become more informative when interpreted in the context of the evolving interaction trajectory of the same information. For example, the same acceptance event may reflect different levels of processing depending on whether the user later revisits that information. Incorporating such process-level evidence could enrich existing models of knowledge gain \cite{roy2020exploring, nasser2024rulkkg}, AI reliance \cite{sivaraman2026intelligent, raees2026people}, and user engagement \cite{abbas2026having, liu2026sensing}.

Our behavioral analyses further identify \textbf{sustained active engagement} as a particularly informative family of traces for modeling uptake. Prior work has linked individual behaviors and broader interaction patterns to user outcomes: frequent copy-paste and repeated LLM reference are associated with overreliance \cite{liu2026behavioral}, while modifying rather than directly accepting AI-generated text has been linked to stronger writing outcomes \cite{yang2025modifying}. Related work also highlights verification and response integration as important forms of critical engagement with AI-generated information \cite{lee2025impact}, and shows that prompting users to reflect can reduce overreliance in AI-assisted decision making \cite{li2026guided}. Our findings show that uptake depends not only on whether users encounter or adopt information, but also on how they continue to engage with that same information afterward. Repeated inspection, revision, Chat--Draft coordination, source checking, and final-stage review all reflect continued engagement with the same information over time. Thus, the same copy-paste or acceptance event may carry different evidence depending on whether the user later revisits, revises, or checks that content. Future user models should therefore treat sustained active engagement with the same information as an important behavioral signal, rather than solely relying on whether an action occurred or how often it occurred.

\subsection{Extending Performance-Based Uptake Assessment to Open-Ended Artifacts}
In our study, we adopt a performance-based measure of information uptake following prior work \cite{collins2016assessing,gadiraju2018analyzing,freed2013effects,pajor2020effects}. For each atomic information unit, uptake is labeled by whether the user correctly answers the corresponding post-task recognition question. This provides a directly scored behavioral outcome that complements subjective judgments, which can be affected by individual calibration and remain misleading when users are confident but incorrect \cite{buccinca2021trust,yin2019understanding}. Existing studies often organize measurement around predefined structural units, such as documents \cite{guo2012beyond}, paragraphs \cite{seifert2017focus}, individual HTML elements \cite{grusky2017modeling}, or predefined content blocks \cite{zhu2024modeling}. However, this creates a gap between lab-based measurement and open-ended real-world tasks, where the final content and assessment units emerge dynamically through interaction. Our approach instead derives atomic assessment targets from each final artifact, preserving the open-ended nature of the task while retaining structured, performance-based assessment. This gives our findings greater real-world relevance and makes them more applicable to actual user workflows.

More broadly, this approach can help researchers bring performance-based assessment into more realistic HCI workflows, such as writing \cite{dang2025corpusstudio}, information seeking \cite{duck2025finding,mei2025interquest}, and planning \cite{abbas2026having}. By deriving assessment targets from what users actually produce during the task, it enables more personalized and diagnostic evaluation of human-AI systems. Prior HCI work has personalized evaluation or support based on user-level proficiency \cite{schoni2025s}, performance on predefined tasks \cite{wang2021providing}, or predicted user states \cite{liu2026sensing}. These approaches adapt scores, feedback, or interventions within a predefined assessment space. Our approach instead makes the assessment targets themselves user-specific by deriving them from the content each user actually produces. This enables content-level diagnosis of which specific information a user has taken up in an open-ended workflow.

In this work, we use immediate recognition as the performance probe, but the same strategy could extend to other uptake-related outcomes. Delayed recognition or recall could assess whether information remains accessible beyond the immediate task \cite{zdanovic2022influence}; comprehension assessments could test whether users understand its meaning \cite{kim2017explaining}; and transfer assessments could examine whether users can apply it in new contexts or tasks \cite{urgo2022learning}. Extending the same artifact-adaptive strategy across these outcomes could provide a broader view of how deeply and durably users take up information.

\subsection{From Specialized Sensing to Low-Cost Interaction Traces in Everyday Workflows}
TraceMind suggests a practical path toward uptake-aware systems by relying on low-cost interaction traces. Mouse, keyboard, and interface-level traces are already produced during ordinary browser and editor use, without requiring dedicated sensing hardware, continuous camera access, or repeated self-report \cite{buscher2009you,hutt2022feasibility}. This makes them promising signals for integrating user-state modeling into everyday co-generation workflows.

Our results suggest that interaction events in dynamic co-generation interfaces should be anchored to both the semantic identity of the information and its screen layout. Prior web interaction work typically ties behavior to page elements or spatial regions: viewport exposure has been used to model attention over individual page elements \cite{grusky2017modeling}, mouse activity has been aligned with content blocks \cite{boi2016reconstructing}, and cursor-based models have incorporated layout-specific features for heterogeneous SERP layouts \cite{villaizan2025adsight}. In these settings, the page element often serves as both the content identifier and the spatial anchor. This coupling breaks in co-generation: the same information may be paraphrased, move from Chat to Draft, and later reflow into a different geometry. TraceMind therefore performs two distinct alignment steps: semantic alignment identifies spans that express the same information across Chat and Draft, while layout alignment recovers where each matched span was rendered over time. Our ablations show that both contribute to prediction. This suggests that interaction modeling in editable, generative interfaces should track semantic identity and rendered geometry separately rather than assume that a stable interface element can represent both.

In addition, our soft-boundary analysis extends the idea of soft bounding regions from prior work to low-cost interaction-trace modeling. Eye-tracking studies commonly expand dynamic areas of interest beyond their exact bounding boxes to avoid missing relevant gaze samples \cite{faraji2023toolkit}, while prior analyses show a trade-off between capturing relevant activity and introducing false associations as AOI margins increase \cite{orquin2016areas}. We observe a similar pattern for low-cost pointer traces: restricting pointer trajectories to the exact target bounding box substantially weakened prediction, whereas retaining a moderate amount of nearby activity provided useful additional evidence. Expanding the region further, however, offered little additional benefit. These results suggest that low-cost interaction modeling should use bounded, target-relative spatial context rather than treat pointer activity as a binary inside-or-outside event.

Finally, information uptake is only one user state that can be modeled from low-cost interaction traces. Prior work has used mouse and interface behavior to infer attention \cite{arapakis2020learning}, engagement \cite{grinberg2018identifying}, and knowledge gain \cite{yu2018predicting}. Our results suggest that combining semantic and layout alignment could extend such modeling to more dynamic tasks and interface settings, helping future HCI systems unobtrusively infer users' evolving states during everyday interaction.

\subsection{Toward Uptake-Aware Human-LLM Co-Generation Systems}
Our findings motivate uptake-aware human--LLM co-generation systems, particularly in consequential settings where users must share, submit, or act on AI-assisted content. Beyond existing support for generating and refining content \cite{liu2024ai,kim2024diarymate,chung2022talebrush}, uptake-aware systems could surface information with weak evidence of uptake for targeted review, making human-LLM collaboration more transparent while preserving user agency and meaningful oversight.

This uptake-aware layer could complement existing HCI support for AI-assisted writing, which has largely focused on suggestions, writing scaffolds, and co-writing workflows \cite{lee2022coauthor,dhillon2024shaping,monge2025investigating}. In accuracy-sensitive settings, such as scientific writing \cite{monge2025investigating} and contract drafting \cite{zeng2025contractmind}, uptake-aware support could prioritize claims with weak evidence of uptake for verification, helping users focus review effort where accuracy matters most. In public-facing communication, such as journalism \cite{nishal2026helping} and online posting \cite{jakesch2023co}, it could surface AI-assisted content with weak evidence of uptake before publication, helping authors remain aware of and in control of what they communicate. Across these settings, uptake awareness complements support for content production by making users' engagement with the resulting information itself an explicit design concern.

This uptake-aware layer could also extend beyond content production to settings where users rely on LLM-provided information. In LLM-assisted learning, where monitoring one's own learning is critical \cite{lee2025learning}, uptake-aware support could infer which information learners have not fully taken up, enabling more targeted and personalized learning support. In AI-assisted decision making, where overreliance on AI remains a concern \cite{li2025text,li2026guided}, uptake-aware signals could infer when users rely on decision-relevant information they have not fully processed, supporting more critical evaluation of AI advice.

At the same time, uptake-aware support should remain soft, inspectable, and privacy-conscious. Because interaction traces provide probabilistic rather than definitive evidence of uptake, predictions should flag content for possible review rather than judge what users know or understand. Systems could consider information importance and interruption cost when deciding whether to intervene \cite{klauck2017noticeable,mehrotra2016my}, and use occasional user feedback to calibrate such support \cite{su2026tell}. Although TraceMind relies only on low-cost interaction traces rather than cameras or other dedicated sensing, these traces may still reveal sensitive content and behavioral patterns. Practical systems should therefore minimize data collection, favor local or on-device processing when feasible \cite{park2025know,jiang2026hear}, and give users control over what data are recorded and retained \cite{alves2026exploring}.

\subsection{Limitations and Future Work}
First, following prior work, we operationalize information uptake as users' immediate recognition of atomic information units in the final draft. This captures recognition-level uptake, but not comprehension, application, or longer-term retention. Future work should examine whether the same interaction traces generalize to these outcomes.
Second, post-task recognition questions were generated and screened through an automated LLM-based pipeline without independent human validation. While this design enabled assessment to adapt to each open-ended artifact, automatically generated items may not achieve psychometric quality and item discrimination comparable to expert-written items \cite{laupichler2024large,holzing2026fine}. Future studies should further combine blinded expert review with empirical item analysis.
Third, our participants were predominantly experienced LLM users in China, and our study covered three representative everyday writing tasks. Future work should evaluate TraceMind across more diverse populations and tasks.
Finally, we prioritized low-cost interaction traces over specialized sensing hardware. In our study, participants used laptops with external mice to ensure consistent capture. Future work should test whether TraceMind transfers to trackpads, touchscreens, and other input modalities.

%% file: data/chap8.tex
\section{Conclusion}
Information uptake is a complex, unfolding process that deserves greater attention to how human-LLM systems are evaluated and designed. By extending performance-based uptake assessment to open-ended artifacts and aligning low-cost interaction traces with atomic information units, our work provides a concrete approach for assessing and modeling uptake during co-generation. We encourage future HCI research to move beyond whether AI-generated content is adopted and toward better ways of assessing and supporting what users actually take up, enabling uptake-aware human-LLM co-generation systems grounded in low-cost interaction traces.

%% file: data/appendix.tex
\clearpage
\newpage

\section{DATA COLLECTION STUDY DETAILS}
This section provides additional details about the data collection study, including participant demographics and the complete participant-facing instructions for the three co-generation tasks.

\subsection{Participant Demographics}
\label{appendix: Participant Demographics}
We provide the participants’ demographics and LLM usage background (see \autoref{tab:participant-demographics}).

\subsection{Participant-Facing Task Instructions}
\label{appendix: Participant-Facing Task Instructions}
We provide the complete participant-facing instructions for the three co-generation tasks. The instructions were originally presented in Chinese and are reproduced here in English. To preserve anonymity, we replaced potentially identifying proper nouns, such as universities, locations, venues, and speaker names, with consistent placeholders while preserving the original task scenarios.

\subsubsection{Club Team-Building Announcement}
\mbox{}\par
\vspace{4pt}
\begin{tcolorbox}[
    breakable,
    colback=gray!8,
    colframe=gray!25,
    boxrule=0.1pt,
    arc=3mm,
    left=4mm,
    right=4mm,
    top=3mm,
    bottom=3mm,
    before skip=6pt,
    after skip=12pt,
]
\small
You are responsible for organizing your club's annual team-building activity, which aims to promote communication and collaboration among members. The following arrangements have been confirmed:
\begin{itemize}
    \setlength{\itemsep}{1pt}
    \setlength{\parskip}{0pt}
    \setlength{\parsep}{0pt}
    \item During the day, the group will travel by coach from University A to Historic Town H and visit Temple T, the South Gate, and the historic district.
    \item In the evening, the group will dine at Restaurant R and try the town's local ``Eight Bowls'' cuisine before returning together by coach.
\end{itemize}

Please use an LLM to help you draft a team-building announcement for all club members. The announcement should introduce the purpose of the activity, the complete itinerary, the attractions, and the restaurant. It should also include practical information that participants may need, such as transportation arrangements, attire suggestions, an introduction to the local cuisine, activity preparation, and other relevant precautions.

Submit a complete team-building announcement that you believe could be sent directly to club members. Treat the final draft as a formal announcement ready for public release and prepare it carefully.
\end{tcolorbox}

\subsubsection{Roommate Recruitment Post}
\mbox{}\par
\vspace{4pt}
\begin{tcolorbox}[
    breakable,
    colback=gray!8,
    colframe=gray!25,
    boxrule=0.1pt,
    arc=3mm,
    left=4mm,
    right=4mm,
    top=3mm,
    bottom=3mm,
    before skip=6pt,
    after skip=12pt,
]
\small
You will attend a two-month summer program at University B and stay in a shared apartment. The following arrangements have been confirmed:
\begin{itemize}
    \setlength{\itemsep}{1pt}
    \setlength{\parskip}{0pt}
    \setlength{\parsep}{0pt}
    \item Apartment A is located in District D.
    \item The unit type is D2--4$\times$4. You and two other students in a joint program between University A and University B have confirmed your stay, and you need to recruit one additional roommate.
\end{itemize}

Please use an LLM to help you draft a recruitment post for potential roommates. You may introduce the apartment, bedrooms, amenities, surrounding neighborhood, transportation, and move-in arrangements to help potential roommates understand the living situation. You should also describe your basic expectations for shared living, such as how expenses will be divided, cleaning, visitors, daily schedules, or other house rules.

Submit a complete roommate recruitment post that you believe could be published publicly. Treat the final draft as a formal post ready for public release and prepare it carefully.
\end{tcolorbox}

\subsubsection{Lecture Announcement}
\mbox{}\par
\vspace{4pt}
\begin{tcolorbox}[
    breakable,
    colback=gray!8,
    colframe=gray!25,
    boxrule=0.1pt,
    arc=3mm,
    left=4mm,
    right=4mm,
    top=3mm,
    bottom=3mm,
    before skip=6pt,
    after skip=12pt,
]
\small
The university will hold an ``AI Lecture Series'' event at 2:00~p.m. the following Friday. Three academicians from University A---Scholar P1, Scholar P2, and Scholar P3---will discuss the current state and future development of artificial intelligence. Each scholar will give an academic talk, followed by a Q\&A session.

Please use an LLM to help you draft an event announcement for all students at the university. Briefly introduce each scholar, including, for example, their research areas and representative work. The announcement should also include information that attendees may need, such as the event theme, the value of attending, suggested preparation, and topics they may wish to follow or ask about.

Submit a complete announcement that you believe could be sent directly to all students at the university. Treat the final draft as a formal announcement ready for public release and prepare it carefully.
\end{tcolorbox}

\section{TRACEMIND FRAMEWORK DETAILS}
\subsection{Workflow-informed Interaction Features}
\label{appendix: Workflow-informed Interaction Features}
We provide the complete set of workflow-informed interaction features used in TraceMind. Table~\ref{tab:workflow-informed-features} lists all 40 features, organized by the eight behavior types.

\onecolumn
{
\scriptsize                          
\renewcommand{\arraystretch}{0.8}
\begin{tabularx}{\textwidth}{c c c c >{\centering\arraybackslash}X c >{\centering\arraybackslash}X >{\centering\arraybackslash}X c}
\caption{Participant demographics and LLM usage background including participant ID, gender, age, education, LLM familiarity, LLM Use Frequency, frequency of LLM use for writing, reliance on LLM, and LLM use experience. LLM familiarity categories: NU = Never used, unfamiliar, BA = Basic awareness (heard of / used once or twice), OU = Occasional use, knows common features, FA = Familiar, can complete common tasks proficiently, VF = Very familiar, can use flexibly or guide others. LLM reliance degree in writing categories: ES = Entirely self-written, MS = Mostly self-conceived and written, CL = Collaborate: discuss ideas or revise with LLM, LF = LLM provides first drafts or frameworks, LD = Almost entirely LLM-dependent.} 
\label{tab:participant-demographics}\\
\toprule
\textbf{PID} &
  \textbf{Gender} &
  \textbf{Age} &
  \textbf{Education} &
  \textbf{LLM Familiarity} &
  \textbf{LLM Use Frequency (past month)} &
  \textbf{LLM Writing Frequency} &
  \textbf{LLM Reliance Degree} &
  \textbf{LLM Use Experience} \\ \midrule
\endfirsthead
\multicolumn{9}{c}{\textit{Continued from previous page}} \\
\toprule
\textbf{PID} & \textbf{Gender} & \textbf{Age} & \textbf{Education} & \textbf{LLM Familiarity} & \textbf{LLM Use Frequency (past month)} & \textbf{LLM Writing Frequency} & \textbf{LLM Reliance Degree} & \textbf{LLM Use Experience} \\
\midrule
\endhead

\bottomrule \\
\endfoot

\endlastfoot
P1  & Male   & 21 & Undergraduate & FA & Multiple times daily  & Always    & LF & 2-3 years          \\
P2  & Male   & 20 & Undergraduate & FA & Multiple times daily  & Always    & LF & 1-2 years          \\
P3  & Female & 22 & Undergraduate & VF & Multiple times daily  & Always    & LD & 2-3 years          \\
P4  & Female & 19 & Undergraduate & VF & Multiple times daily  & Always    & LD & 2-3 years          \\
P5  & Female & 19 & Undergraduate & FA & Multiple times daily  & Always    & LD & 1-2 years          \\
P6  & Male   & 23 & Doctoral      & VF & Multiple times daily  & Always    & CL & 2-3 years          \\
P7  & Female & 23 & Master's      & FA & A few times per week  & Sometimes & CL & 2-3 years          \\
P8  & Male   & 26 & Master's      & FA & Multiple times daily  & Always    & LF & 1-2 years          \\
P9  & Female & 21 & Undergraduate & FA & A few times per week  & Often     & CL & 2-3 years          \\
P10 & Male   & 20 & Undergraduate & FA & A few times per week  & Often     & LF & 2-3 years          \\
P11 & Female & 21 & Undergraduate & FA & Almost daily          & Often     & CL & 1-2 years          \\
P12 & Female & 18 & Undergraduate & FA & Almost daily          & Often     & LF & 2-3 years          \\
P13 & Male   & 22 & Master's      & FA & Multiple times daily  & Always    & LD & 1-2 years          \\
P14 & Male   & 19 & Undergraduate & FA & A few times per week  & Often     & LD & 7-12 months        \\
P15 & Male   & 21 & Undergraduate & FA & Almost daily          & Often     & CL & 1-2 years          \\
P16 & Male   & 22 & Undergraduate & FA & Almost daily          & Sometimes & CL & 2-3 years          \\
P17 & Female & 19 & Undergraduate & FA & Almost daily          & Often     & CL & 1-2 years          \\
P18 & Female & 20 & Undergraduate & FA & A few times per month & Always    & LD & 7-12 months        \\
P19 & Female & 20 & Undergraduate & FA & Almost daily          & Often     & CL & 1-2 years          \\
P20 & Male   & 22 & Master's      & VF & Multiple times daily  & Always    & CL & More than 3 years  \\
P21 & Male   & 24 & Master's      & OU & Almost daily          & Often     & CL & 1-2 years          \\
P22 & Female & 23 & Master's      & VF & Multiple times daily  & Always    & LD & 2-3 years          \\
P23 & Male   & 22 & Undergraduate & OU & A few times per month & Often     & LF & 2-3 years          \\
P24 & Female & 22 & Master's      & VF & Multiple times daily  & Always    & LD & More than 3 years  \\
P25 & Female & 30 & Undergraduate & OU & Almost daily          & Sometimes & LF & 1-2 years          \\
P26 & Male   & 21 & Undergraduate & FA & Almost daily          & Always    & CL & 2-3 years          \\
P27 & Male   & 23 & Master's      & FA & Almost daily          & Sometimes & CL & 1-2 years          \\
P28 & Male   & 21 & Undergraduate & FA & Almost daily          & Often     & CL & 1-2 years          \\
P29 & Female & 21 & Undergraduate & FA & Multiple times daily  & Always    & LD & 1-2 years          \\
P30 & Female & 23 & Undergraduate & FA & A few times per week  & Often     & CL & 1-2 years          \\
P31 & Female & 23 & Master's      & FA & Multiple times daily  & Always    & LD & 2-3 years          \\
P32 & Female & 23 & Master's      & FA & Almost daily          & Sometimes & LF & 2-3 years          \\
P33 & Male   & 21 & Undergraduate & FA & Almost daily          & Always    & LD & More than 3 years  \\
P34 & Female & 24 & Master's      & FA & Multiple times daily  & Always    & LF & More than 3 years  \\
P35 & Male   & 20 & Undergraduate & FA & A few times per week  & Sometimes & LF & 1-2 years          \\
P36 & Female & 23 & Undergraduate & FA & A few times per week  & Always    & LD & 2-3 years          \\
P37 & Male   & 20 & Undergraduate & FA & Almost daily          & Sometimes & CL & 1-2 years          \\
P38 & Female & 23 & Undergraduate & VF & Almost daily          & Always    & LF & 2-3 years          \\
P39 & Male   & 21 & Undergraduate & VF & Almost daily          & Often     & LF & More than 3 years  \\
P40 & Male   & 23 & Undergraduate & VF & Multiple times daily  & Often     & CL & 2-3 years          \\
P41 & Female & 20 & Undergraduate & FA & Almost daily          & Often     & LD & 1-2 years          \\
P42 & Male   & 20 & Undergraduate & FA & A few times per week  & Sometimes & MS & 1-2 years          \\
P43 & Female & 20 & Undergraduate & FA & Almost daily          & Often     & LF & 2-3 years          \\
P44 & Male   & 20 & Undergraduate & FA & Multiple times daily  & Always    & LD & 2-3 years          \\
P45 & Female & 22 & Undergraduate & FA & Multiple times daily  & Often     & LF & 2-3 years          \\
P46 & Male   & 23 & Undergraduate & FA & Multiple times daily  & Often     & LF & 1-2 years          \\
P47 & Male   & 20 & Undergraduate & FA & A few times per week  & Often     & CL & Less than 6 months \\
P48 & Male   & 21 & Undergraduate & FA & Multiple times daily  & Often     & LF & 1-2 years          \\
P49 & Female & 21 & Undergraduate & FA & Multiple times daily  & Always    & LD & 2-3 years          \\
P50 & Female & 19 & Undergraduate & FA & Almost daily          & Often     & CL & 7-12 months        \\
P51 & Male   & 20 & Undergraduate & FA & A few times per week  & Sometimes & LF & 7-12 months        \\
P52 & Male   & 21 & Undergraduate & FA & Multiple times daily  & Often     & CL & 1-2 years          \\
P53 & Female & 20 & Undergraduate & FA & Multiple times daily  & Often     & CL & 2-3 years          \\
P54 & Male   & 22 & Master's      & FA & Multiple times daily  & Always    & LD & 1-2 years          \\
P55 & Male   & 21 & Undergraduate & FA & Almost daily          & Always    & LD & 1-2 years          \\
P56 & Female & 20 & Undergraduate & FA & Multiple times daily  & Often     & CL & 1-2 years          \\
P57 & Female & 22 & Undergraduate & FA & Multiple times daily  & Often     & CL & 1-2 years          \\
P58 & Male   & 26 & Master's      & FA & A few times per week  & Often     & CL & 1-2 years          \\
P59 & Female & 21 & Undergraduate & FA & Almost daily          & Often     & LF & 1-2 years          \\
P60 & Female & 23 & Undergraduate & FA & Multiple times daily  & Often     & LD & 2-3 years          \\
P61 & Male   & 20 & Undergraduate & FA & Almost daily          & Often     & LF & 7-12 months        \\
P62 & Male   & 21 & Undergraduate & FA & Multiple times daily  & Often     & LF & 1-2 years          \\ \bottomrule
\end{tabularx}
}

\begin{table*}[htbp]
\centering
\small
\caption{Workflow-informed operation features used in TraceMind.}
\label{tab:workflow-informed-features}
\begin{tabular}{p{0.25\linewidth} p{0.65\linewidth}}
\toprule
\multicolumn{1}{c}{\textbf{Behavior type}} &
\multicolumn{1}{c}{\textbf{Operationalized features}} \\
\midrule

Chat-side repeated inspection
&
number of Chat visits, number of Chat revisits, number of delayed Chat revisits, binary indicator of delayed Chat revisit, maximum time gap between Chat revisits, normalized Chat visual revisit count per occurrence \\

\midrule

Draft-side repeated inspection
&
number of Draft visits, number of Draft revisits, binary indicator of Draft revisit, number of delayed Draft revisits, binary indicator of delayed Draft revisit, mean time gap between Draft revisits, maximum time gap between Draft revisits, active Draft visit rate, normalized Draft visual revisit count per information-unit member \\

\midrule

Draft-side inspection after editing
&
number of Draft revisits after editing, binary indicator of Draft revisit after editing\\

\midrule

Draft editing and revision
&
number of manual edits, number of edit bouts, binary indicator of any edit, edits per Draft visit, normalized manually inserted text length \\

\midrule

Copy-and-paste-based transfer
&
normalized Draft paste count per occurrence, normalized Draft paste count per information-unit member, normalized pasted text length per information-unit member, normalized Draft copy count per occurrence, normalized verified copy-paste chain count per occurrence, minimum latency from Chat copy to Draft paste \\

\midrule

Cross-surface text selection
&
normalized Chat text selection count per occurrence, normalized Draft text selection count per occurrence \\

\midrule

Cross-surface coordination
&
number of cross-surface switches, number of cross-surface engagement events, number of cross-surface bouts, number of distinct visited surfaces, number of distinct visited occurrences, number of cross-occurrence switches, number of event-linked opportunities \\

\midrule

Source checking before revision
&
binary indicator of source check before revision, number of source checks before revision, minimum latency from source check to subsequent revision \\

\bottomrule
\end{tabular}
\end{table*}